\documentclass[iop]{emulateapj}
\usepackage{graphicx}
\usepackage{epstopdf}

\slugcomment{Accepted for publication in the Astrophysical Journal}

\def\etal{\hbox{et al.}$\,$}
\def\OII{\hbox{[O\,II]}$\,\,$}
\def\Hd{\hbox{H$\delta$}$\,\,$}
\def\Ha{\hbox{H$\alpha$}$\,$}
\def\Hb{\hbox{H$\beta$}$\,$}

\def\EWOII{\hbox{EW([O II])}$\,$}
\def\EWHd{\hbox{EW(H$\delta$)}$\,$}
\def\DHd{\hbox{$\Delta\EWHd$}$\,$}
\def\24m{\hbox{24\,$\micron$}$\,$}
\def\IRAS25{\hbox{25\,$\micron$}$\,$}
\def\tightminus{\!\!-\!\!}
\newcommand{\microJy}{$\mu$Jy}

\begin{document}

\shortauthors{Oemler et al.}
\shorttitle{IMACS Cluster Building Survey. I}

\title{The IMACS Cluster Building Survey. I. Description of the Survey and Analysis Methods\footnotemark[1]}

\footnotetext[1]{This paper includes data gathered with the 6.5 meter Magellan Telescopes located at Las Campanas Observatory, Chile}

\author{Augustus Oemler, Jr.\altaffilmark{2}, Alan Dressler\altaffilmark{2}, 
Michael G. Gladders\altaffilmark{3}, Jane R.\ Rigby\altaffilmark{4}, Lei Bai\altaffilmark{5}, Daniel Kelson\altaffilmark{2}, Edward Villanueva\altaffilmark{2}, Jacopo Fritz\altaffilmark{6},  George Rieke\altaffilmark{7}, Bianca M.\ Poggianti\altaffilmark{8}, \& Benedetta Vulcani\altaffilmark{8}\altaffilmark{9}}

\altaffiltext{2}{The Observatories of the Carnegie Institution for Science, 813 Santa Barbara St., Pasadena, California 91101-1292, USA, oemler@obs.carnegiescience.edu}
\altaffiltext{3}{Department of Astronomy \& Astrophysics, University of Chicago, Chicago, IL 60637, USA} 
\altaffiltext{4}{Observational Cosmology Lab, NASA Goddard Space Flight Center, Greenbelt, MD 20771, USA}
\altaffiltext{5}{Department of Astronomy and Astrophysics, University of Toronto, 50 St. George Street, Toronto, ON M5S 3H4, Canada}
\altaffiltext{6}{Sterrenkundig Observatorium, Universiteit Gent, Krijgslaan 281 S9, B-9000 Gent, Belgium}
\altaffiltext{7}{Steward Observatory, University of Arizona, Tucson, AZ 8572}
\altaffiltext{8}{INAF-Osservatorio Astronomico di Padova, vicolo 
dell'Osservatorio 5, 35122 Padova, Italy}
\altaffiltext{9}{\,Department of Astronomy, Padova University, Vicolo Osservatorio 3, I-35122 Padova, Italy}

\begin{abstract}
The IMACS Cluster Building Survey uses the wide field spectroscopic capabilities of the IMACS spectrograph on the 6.5m Baade Telescope to survey the large--scale environment surrounding rich intermediate--redshift clusters of galaxies. The goal is to understand the processes which may be transforming star--forming field galaxies into quiescent cluster members as groups and individual galaxies fall into the cluster from the surrounding supercluster. This first paper describes the survey: the data taking and reduction methods. We provide new calibrations of star formation rates derived from optical and infrared spectroscopy and photometry. We demonstrate that there is a tight relation between the observed star formation rate per unit B luminosity, and the ratio of the extinctions of the stellar continuum and the optical emission lines. With this, we can obtain accurate extinction--corrected colors of galaxies. Using these colors as well as other spectral measures, we determine new criteria for the existence of ongoing and recent starbursts in galaxies. 
\end{abstract}


\section{Introduction}

That the properties of galaxies differ with environment has been recognized at least since Hubble (1936). That much of the difference between clusters and the field is of recent origin has been known since Butcher \& Oemler (1978).  However the mechanisms that have produced these population differences are still in dispute. Many processes have been suggested which are capable of transforming field--like populations into the predominantly early Hubble types seen in clusters today. These include gas stripping by galaxy--galaxy collisions (Spitzer \& Baade 1951), gas stripping by ram pressure from the intracluster medium (Gunn \& Gott  1972), a shutoff in gas replenishment (Larson, Tinsley, \& Caldwell 1980),  tidal shocks, either due to the cluster core (Byrd \& Valtonen 1990,  Henriksen \& Byrd 1996), to unvirialized subclusters (Gnedin 2003), or to other galaxies (Richstone \& Malmuth 1983, Icke 1985, Moore \etal 1996), and galaxy-galaxy mergers (Dressler \etal 1999, van Dokkum \etal 1999). Since all these processes produce, by design, the same outcome, and have, again by design, a qualitatively similar dependence on environment, distinguishing between them is challenging at best.  Given that most extant observations consist of  snapshots at one epoch of the cores of individual clusters, it is hardly surprising that the responsible process(es) have still not been unambiguously identified.

The IMACS Cluster Building Survey is an attempt to resolve this problem by following the evolution of galaxies as they  move from the supercluster environment, through the infall stage and end finally with  incorporation into the virialized central cluster. It takes advantage of the very wide field and high multiplexing of the IMACS spectrograph on the Baade Telescope (Dressler \etal 2011), which allow one to observe, in one exposure, hundreds of  galaxies over a 30\arcmin\  field, equivalent to a circle with radius of about 5 Mpc surrounding a cluster at redshift 0.4. The goal is to identify the changes that occur in galaxies as they move from field--like environments into increasingly dense and massive assemblies. The much longer baselines of time and environment which these data provide should help distinguish between the various candidate processes for transforming galaxies.

In this paper we describe the design and execution of the survey, up through the production of redshifts, luminosities, colors, masses, and star formation rates of galaxies. We describe new calibrations of star formation rates based on optical and infrared photometry and spectroscopy and we discuss several methods for detecting starbursts using the available data.  The immediately following papers (Dressler \etal 2013- Paper II,  Oemler \etal 2013- Paper III, Gladders \etal 2013- Paper IV) discuss certain aspects of the behavior of galaxies in the cluster and supercluster environment, and the evolution of the field galaxy population. Two papers using ICBS data to analyze environmental variations in the mass function of galaxies have already been completed (Vulcani \etal 2012, 2013). Future papers will address other aspects of both the field and cluster populations. Throughout this and following papers we shall assume cosmological parameters of $H_o=71$, $\Omega_{matter}=0.27$, $\Omega_{tot}=1.0$. 

\section{Observations}

We wish to map a group of rich intermediate redshift clusters including still-forming objects unlikely to be discovered by x-ray searches. To do this we use the cluster red-sequence detection method (Gladders \& Yee 2000), applied to data from the Red-Sequence Cluster Survey (RCS; Gladders \& Yee 2005) and the Sloan Digital Sky Survey (SDSS, York \etal 2000) Data Release 2 (DR2). The RCS data are $R_c$ and $z'$ imaging to a depth sufficient to detect clusters to $z\sim1.4$, more than sufficient to find clusters at the redshift considered here. The SDSS data are much shallower, but sufficient to find rich/massive clusters at the redshift of interest.

Approximately 50 square degrees of the RCS imaging are readily visible from the Las Campanas Observatory and we initially searched this area for candidate rich clusters in the desired 0.3$<z<$0.5 redshift range. Clusters were detected as over-densities on the sky and in color and magnitude space. The richest systems were considered as candidates for the ICBS program. The lack of sufficient RCS imaging area visible from the southern hemisphere at $6^h$ to $9^h$ and $14^h$ to $20^h$ forced us to use the SDSS DR2 data as a secondary source for rich cluster candidates. In order to ensure a reasonable match in mass between the two cluster sub-samples, we restricted our attention in the SDSS to areas of low-extinction sky comparable in size to the RCS search area, at two RAs selected to facilitate the extensive ICBS data collection. The total comoving volume covered by this search is equal to that in the two Galactic caps out to a redshift $z \sim 0.055$, and should, therefore, include a number of rich clusters comparable to those found in local surveys.
We identified 5 fields which, from the RCS or from our analysis of the SDSS data, appeared to contain very rich clusters at redshifts between 0.3 and 0.5. Preliminary spectroscopy showed that the cluster in one of the five fields was not sufficiently rich to be interesting, leaving 4 fields, two from the RCS and 2 from the SDSS search, whose locations are summarized in Table 1.

\subsection{Spectroscopy}
\subsubsection{Observations}
Slit masks for each field were populated with objects from the photometric catalogs, with a slight preference for brighter objects and a strong preference for objects brighter than $r = 22.5$. After the first mask in each field, later masks contained unobserved objects plus a number of already observed objects, most with poor spectra plus some with adequate spectra to use for repeatability tests. Most (39 slit masks) spectroscopy was done with the full 30\arcmin\  field of the IMACS $f/2$ camera  on the Baade Telescope. However, in order to increase the fraction of objects observed in the cluster cores, where slit overlap issues make it particularly difficult to obtain adequate sampling, 3 masks were obtained using  the GISMO image slicer (q.v. Dressler \etal 2011) on IMACS, and 16 masks were obtained using the LDSS3 spectrograph on the Clay Telescope, which has an 8.3\arcmin\  field. Observations were done with a mixture of stare mode and nod--and--shuffle mode. Typical total integration time per mask were in the range of 3 to 4 hours, divided into individual integrations of 30 to 45 minutes. All IMACS and LDSS3 spectroscopy, except for the LDP observations discussed below, were made at a dispersion of about 2\AA\  per pixel, resulting in a spectral resolution of about 6\AA. Typical image quality during the spectroscopic and imaging observations was 0.6\arcsec--0.8\arcsec.

%Table 1
\begin{deluxetable}{lrr}[h]
\tablecaption{Intermediate Redshift Cluster Fields}
\tablewidth{0pt}
\tablehead{\colhead{Cluster} & \colhead{$\alpha$} & \colhead{$\delta$} }
\startdata
RCS0221 & $2^h 21^m 48^s$ & -03\arcdeg 46\arcmin  \\
RCS1102 & $11^h 02^m 36^s$& -04\arcdeg 40\arcmin  \\
SDSS0845 &$8^h 45^m 30^s$ &  +03\arcdeg 27\arcmin\\
SDSS1500 & $15^h 00^m 30^s $& +01\arcdeg 53\arcmin   \\
\enddata
\end{deluxetable}

The first few slit masks in each field were observed with no band limiting filters, providing spectral coverage between 4300\AA\  and 9300\AA. Later observations were done with a filter limiting coverage to 4800\AA\  to 7800\AA. The shorter spectra allowed more spectra- about 300-  to be packed onto a single mask, but lost coverage of the \Ha line at redshifts $z > 0.19$. In order to obtain \Ha observations of as many cluster galaxies as possible, one mask per field was devoted to observations of  already discovered cluster members through a 1000\AA\  wide filter centered on \Ha at the cluster redshift. 

One mask each was obtained in all fields except SDSS1500 with the IMACS Low Dispersion Prism (LDP).  With a mean resolution $\lambda/\Delta\lambda \sim 30$, the LDP spectra cannot be used to measure individual line strengths, but are sufficient to obtain redshifts with a mean accuracy of about 0.01. Because the spectra are very short, of order 1000 objects can be observed on one slit mask, to a depth considerably fainter than with grism spectroscopy. More information about the LDP prism can be found in Kelson \etal (2012).

The data sets derived from all of these observations are referred to in the following as the ICBS data sets.
 For calibration purposes, we also construct a local sample of galaxies with optical spectroscopy, and optical and \24m infrared photometry. This sample consists of galaxies with SDSS redshifts between 0.04 and 0.08 located in the three fields of the SWIRE survey (Lonsdale \etal 2003) which are within the SDSS survey area. We take \24m flux from the SWIRE observations, and take spectra, redshifts and optical photometry from the SDSS database.

\subsubsection{Spectral Reductions}

With the exception of LDP observations, all IMACS and LDSS3 spectra were reduced using the COSMOS data package (Oemler \etal 2011). In general, all observations of an individual mask, whether from one night or from several observing runs were combined into one  stack of 2--dimensional (flux vs wavelength and slit position) spectral images of individual slits. Using the interactive spectral analysis program {\sc{viewspectra}} from the COSMOS package, a boxcar 1--dimensional extraction of each spectrum was made. Most spectra were extracted over a 1\arcsec\  length along the slit, but wider extractions were used for particularly diffuse galaxies. LDP spectra were reduced using the methods described in Kelson \etal (2012).

Extracted 1--dimensional slit spectra were turned into final calibrated spectra in a 3--step process. Firstly, spectra were put onto a relative flux scale by correcting for the instrumental response using observations of spectrophotometric standards. Because the spectral response of IMACS is very stable, a mean spectral sensitivity function for each instrumental setup was derived by combining all standard star spectra taken with this setup. Secondly, spectra were corrected for atmospheric absorption using the sum of all of the spectra on each slit mask obtained on a given night. Because the galaxies observed  on most  individual slit masks have a wide range of redshift and spectral type, the spectral features of the individual galaxies are averaged out in this summed spectra, leaving only the spectrum of the atmospheric absorption plus a slowly--varying sum of the individual (redshifted) continuum shapes. This spectrum, after removing the continuum variations, was divided into each galaxy spectrum to remove the atmospheric absorption.   Absorption corrections for spectra observed through one of the \Ha filters of cluster members  could not be performed in this manner, since the wavelength range was too narrow and because all of the objects had similar redshifts. Absorption corrections for these masks was performed using sky absorption spectra obtained from other observations of the same cluster. Thirdly, the spectra are put on an absolute flux scale by scaling their values using the difference between the synthetic r magnitude calculated from each spectrum and the total r magnitude of the galaxy. Assuming that there are no significant color gradients between the roughly 1\arcsec\  square area of the galaxy observed with IMACS and the total galaxy, this corrected spectrum represents the total flux of the galaxy. We shall test this assumption later. After calibration, spectra of  galaxies obtained from multiple slit masks  were combined. We have found that better results were obtained if spectra were combined using a single weight for all data points in spectra from each slit mask, rather than using the pixel--to--pixel statistical weights obtained as part of the reduction process. Relative weights varied by a factor of a few between the best and the worst slit masks. The final quality of the spectra is, of course, quite variable. A small fraction (typically  20\%) were too poor to yield a redshift. For the remaining, the median signal--to--noise ratio, per 2\AA\  pixel, at a rest wavelength of 4500\AA, ranges from 30 at $r=19.0$ to 7.5 at $r=22.0$.

Redshifts of galaxies were measured using the cross-correlation method of Kelson  \etal (2005). Analysis of repeat observations of galaxies show a typical error $\sigma(z) \sim 2\times 10^{-4}$. This error is, to first order, independent of spectral type (absorption or emission lines) or galaxy brightness. It is larger than that expected due to the typical wavelength errors determined by measuring atmospheric emission lines $\sigma(z) \sim 6\times 10^{-5}$, but is comparable to that expected due to slit errors. The location of spectra along the slits show a scatter about the expected position of about 1 pixel, equivalent to astrometric errors in the galaxy catalogs of $0.2\arcsec$. A scatter of 1 pixel in the centroid of the galaxies in the dispersion direction corresponds to a redshift error $\sigma(z)$ of $3\times10^{-4}$.

%Figure 1
\begin{figure}[h]
\figurenum{1}
\epsscale{1.1}
\vspace{0.4in}
\plotone{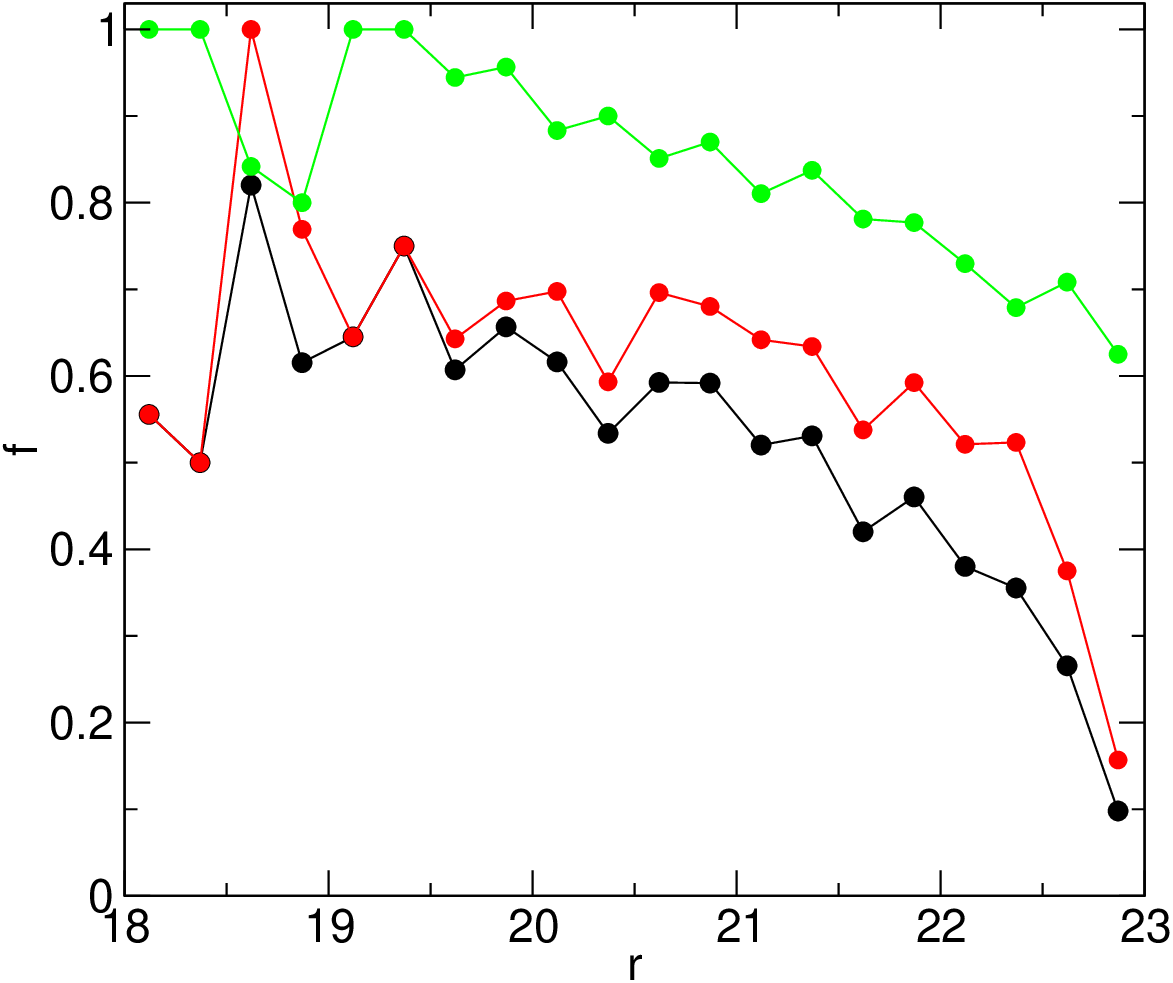}
\figcaption{ Completeness of the RCS1102 sample. black line- fraction of catalog with redshifts; red line- fraction of the catalog that was observed; green line- fraction of observed objects that yielded a redshift.}
\end{figure}

The completeness of the RCS1102 redshift sample, typical of all fields, is presented in Figure 1. The total completeness, shown by the black curve is the product of two factors: the fraction of objects observed, shown by the red curve, and the rate of success in obtaining redshifts from the spectra, shown by the green curve. Although the success rate declines from almost 100\% for the brightest objects to about 75\% by $r=22$, most of the incompleteness is due to object selection, which varies both with magnitude and with position with the field.  The decline with magnitude in the fraction of objects observed, from about 70\% for bright objects to 55\% at $r=22$, is completely due to the algorithm which selects objects to be included on a slit mask, and which has a mild preference for brighter objects. The spatial variation in completeness is small. Because of the number of masks used in each cluster, and because a concerted effort was made to sample well the cluster core, there is very little under-sampling of objects in groups and the cluster. Typically, what under-sampled regions exist lie near the periphery of the field.

Since it is generally easier to identify redshifts for galaxies with strong emission lines, we have checked for such a bias in our redshift catalog. To minimize evolutionary effects (fainter galaxies tend to be at higher redshift, and star formation rates increase with redshift- q.v. Paper III), we examine the fraction of galaxies with $EW([OII]) \ge 5\AA$ in the redshift range $0.15 \le z \le 0.35$. Brighter than $r=20.5$ absorption spectra increase in frequency, because early--type galaxies dominate the bright end of the luminosity function. However, between $r=20.5$ and $r=23.0$, the fraction of emission line galaxies is constant, demonstrating that any spectral type bias is minimal in our sample.

Emission and absorption lines in both the ICBS and SWIRE samples were measured using {\sc{viewspectra}}. This is a semi--automatic process, in which gaussian profiles are fit to each line. The wavelength intervals of continuum side bands, and of the line itself, are specified in advance. However, one can interact with the fitting process to correct for less than perfect fits by altering any of the fitting parameters. Output of the fitting includes equivalent widths of emission and absorption lines as well as fluxes of emission lines. Because the spectra have previously been given an absolute calibration, the resulting fluxes represent the total line flux for the entire galaxy.

In fitting Balmer emission lines, an attempt was made to properly set the continuum in the bottom of the stellar absorption line; however, given the typical signal--to--noise ratios of these spectra, there is a limit to how well this can be done. When measuring the \Hd absorption line, no attempt was made to remove contamination from \Hd emission. Because the \Hd line is often of marginal signal--to--noise ratio in these spectra, an improved measure of EW(\Hd) was constructed by combining the \Hd strength with that of the $H\epsilon$ line, which lies on top of the Ca H line.  $H\epsilon$ was determined from the difference between the Ca K line and the Ca H +$H\epsilon$ equivalent width. Empirically it was found that the \Hd equivalent width is related to that of H+$H\epsilon$ and K as:

\begin{equation}
\begin{array}{rl}
H\tightminus K < 3\AA &  H\delta = 1.73 + 0.5(H\tightminus K) +0.04(H\tightminus K)^2 \\
H\tightminus K\ge3\AA &   H\delta = 0.77(H\tightminus K) +1.35 
\end{array}
\end{equation}

where $H\tightminus K$ is the difference, in Angstroms, between the equivalent withs of H+$H\epsilon$ and K. Our final \Hd values are the average of \Hd and that derived from H and K.

%Figure 2
\begin{figure}[h]
\vspace{0.3in}
\figurenum{2}
\epsscale{1.1}
\plotone{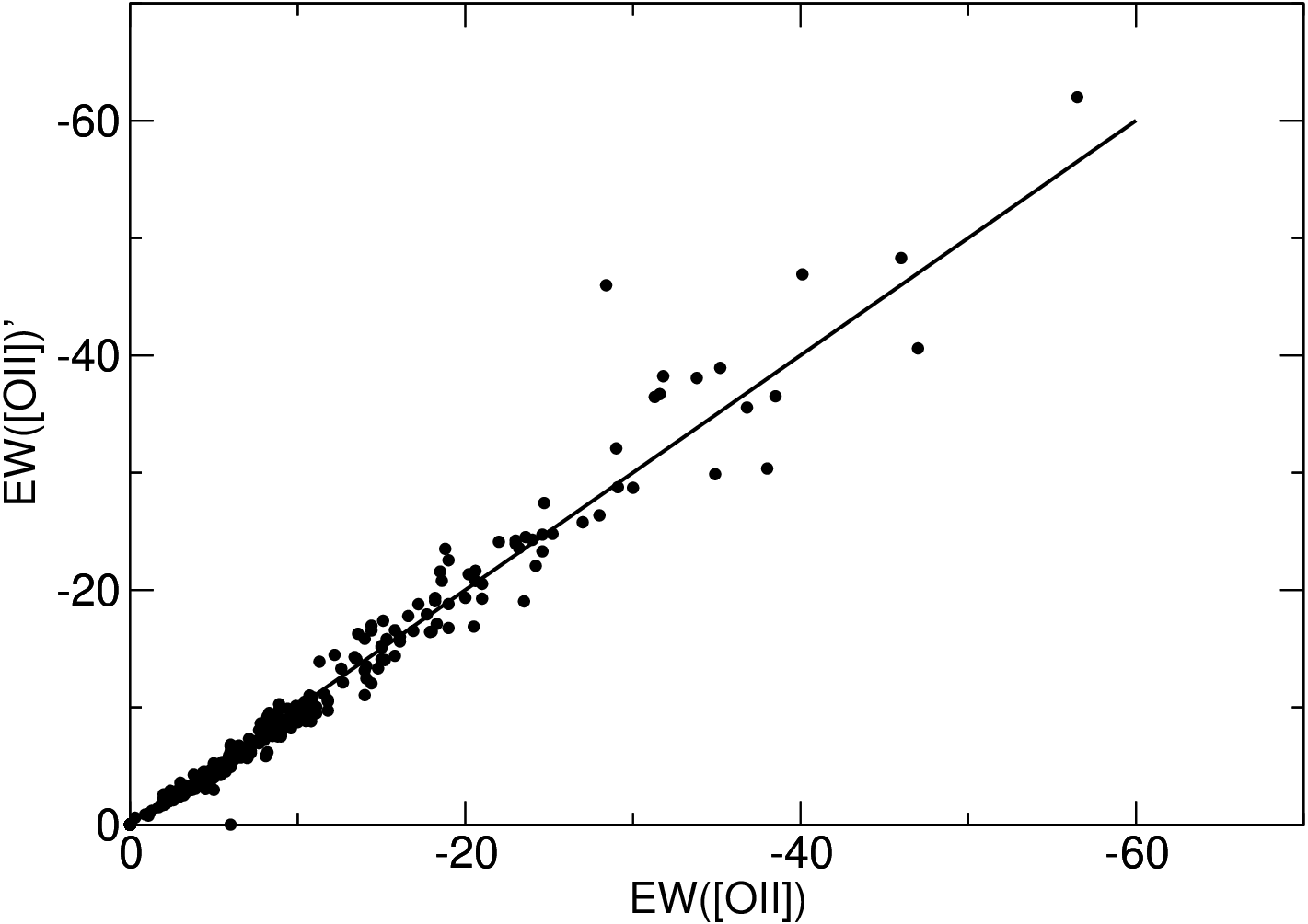}
\figcaption{ Synthetic EW([OII]), as defined by Equations 2 and 3, versus directly--measured EW([OII]), for objects in the SWIRE sample.}
\end{figure}

The [OII] line lies in a rather clean spectral region and is easy to measure except for one complication. Some galaxies, even some with very strong [OII], have a very weak stellar continuum at 3727\AA. Slightly incorrect continuum levels in the reduced spectra due, for example, to sky subtraction errors,  can result in quite large fractional changes in the continuum value, and therefore quite large errors in the equivalent width of [OII]. To avoid this problem, we define a pseudo--continuum at 3700\AA\  based on an empirical relation between 3700\AA\  flux and a combination of M(B) magnitude and B-V color, derived from synthetic photometry of the SWIRE sample of SDSS spectra,  and from this plus measured [OII] flux, determine EW([OII]). The relation between synthetic and directly--measured EW([OII]) is presented, for the SWIRE sample, in Figure 2; a best fit to this relation is:

\begin{equation}
EW([OII]) = -1.13\times10^{-32} L(OII)/10^{-0.4M37}
\end{equation}

where the 3700\AA\  monochromatic magnitude $M_{37}$ is approximated as

\begin{equation}
M_{37} = M_B+0.20+0.67(B\!-\!V-0.09)+0.0359(B\!-\!V-0.09)^2
\end{equation}

Analysis of repeat measurements of individual galaxies show that \OII and \Hb determinations have typical errors of 0.07 dex, and \Hd determinations have typical errors of 0.13 dex, however these errors are quite dependent on spectral quality.

%Figure 3
\begin{figure}[h]
\figurenum{3}
\vspace{0.15in}
\epsscale{0.75}
\plotone{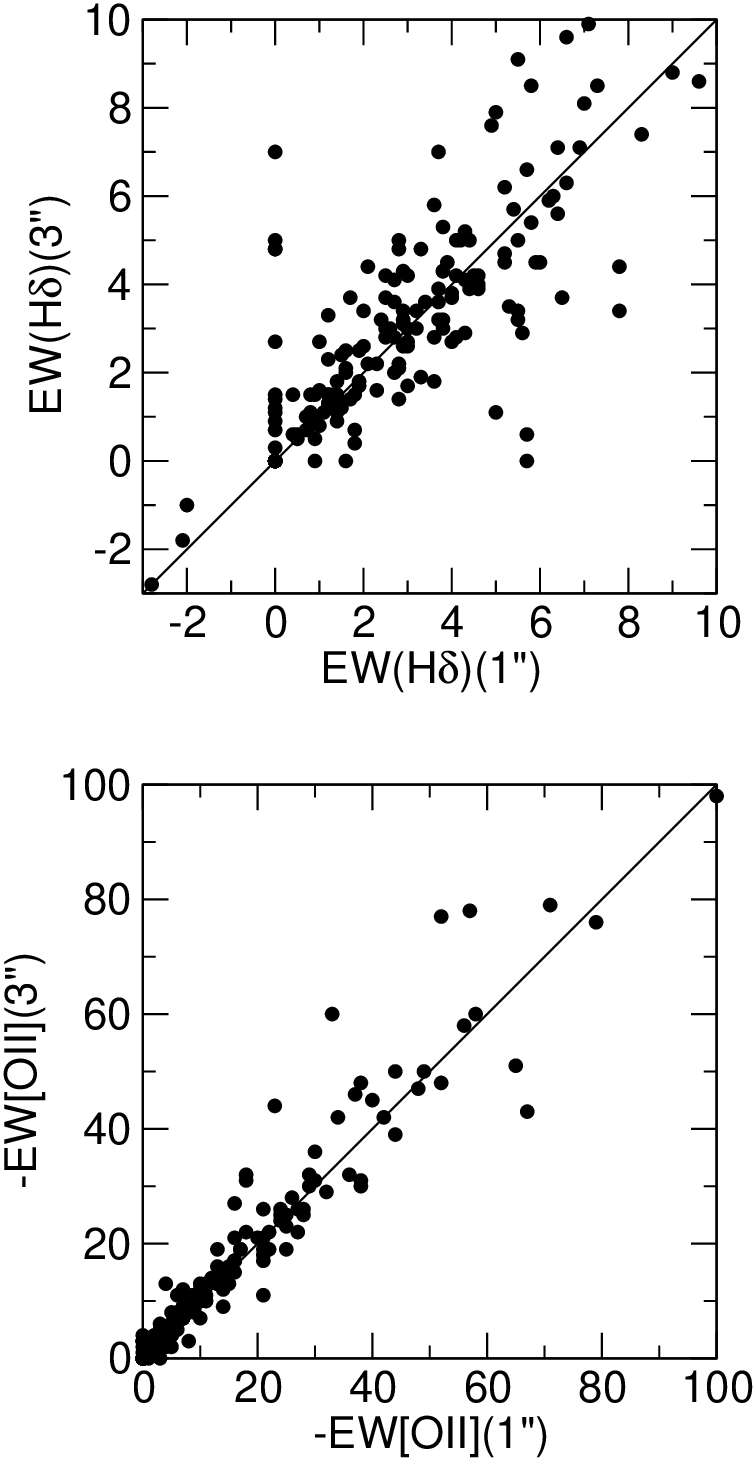}
\figcaption{Synthetic 3\arcsec\  line strengths vs measured 1\arcsec\  line strengths, for galaxies in several masks.}
\end{figure}

The ICBS spectra are typically extracted from an aperture 1\arcsec\  square, equivalent to a 5.3 kpc square area in a $z=0.40$ galaxy. This is considerably smaller than the area containing the bulk of the stars of the typical luminous galaxy. Since  many of the galaxy parameters which we derive depend on an extrapolation from the spectroscopically observed area to the total galaxy, any systematic shift between line strengths in the galactic center to those of the entire galaxy could lead to systematic errors in, for example, star formation rates and internal extinction. To test for this, we take 2-dimensional spectra from several of the best masks and construct synthetic spectra of an area 3\arcsec\  in diameter. Figure 3 compares 1\arcsec\  and 3\arcsec\  measures of EW([OII]) and EW(\Hd). The synthetic spectra are necessarily quite noisy, resulting in considerable scatter, but it is clear that there in no systematic trend of either with area, indicating that our measured line strengths are reliable indicators of the total galaxy values.
\subsection{Photometry}

Direct imaging in the griz bands was obtained for the RCS0221 and SDSS0845 fields with the $f/2$ camera of IMACS. Imaging in the BVRI bands was obtained for RCS1102 and SDSS1500 using the Wide Field CCD camera on the du Pont Telescope. In addition, very deep r--band photometry, complete to $r=25.0$ was obtained for all fields with IMACS. Photometry of the images was performed using SExTractor (Bertin \& Arnauts 1996). Rest frame B-V colors and absolute B magnitudes were derived from the IMACS and du Pont photometry, using k--corrections derived from spectral templates constructed from the spectra of SDSS galaxies in the SWIRE  sample described above. Typical color errors at (20.0, 22.0) mag are (0.007,0.04) in (r-i), (0.015, 0.07) in (i-z), (0.05, 0.10) in (B-V), and (0.04, 0.08) in (V-R). For a small number of galaxies with no direct imaging, we have constructed synthetic rest frame colors and absolute magnitudes from the (fluxed) spectra.

The RCS0221 and SDSS0845 fields were mapped by the Spitzer Space Telescope (Werner et al. 2004) with the MIPS instrument (Rieke et al 2004) in the 24 micron band.  Data were taken in guest observer program 40387, PI Dressler.  Including overheads, the observations lasted 11.7 hr per field. For each cluster, the circular 27\arcmin\ diameter IMACS field of view was tiled with four overlapping MIPS raster-map photometry sequences. This covered the area more efficiently than would scan mapping.  For almost all the IMAC field of view, at least two raster-maps overlap, providing $\ge 980sec$ exposure time per pixel for SDSS0845, and $\ge1069sec$ for RCS0221.

The raw MIPS images were reduced and mosaicked using the MIPS Instrument Team Data Analysis Tool (DAT) (Gordon et al. 2005).   The temporally-varying ecliptic foreground was subtracted separately from each pointing.  Photometry at 24 micron was obtained by fitting the point spread function (PSF), using the IRAF implementation of the DAOPHOT task allstar (Stetson 1987).  The PSF was created empirically from stars in each cluster, and an aperture correction was applied as in Rigby et al. (2008).

An extremely bright foreground carbon star at $\alpha,\delta$ = 08:45:22, +03:27:09 (J2000) contaminates an area $\sim 2.5$\arcmin\ in radius in the cluster SDSS0845.  The SDSS0845 catalogs were edited by hand to remove artifacts caused by this star and account for missing  survey area obscured by the star. Comparing object counts in our fields with the deep \24m counts by Papovich \etal (2004) we determine that our photometry is complete to about 60\microJy, equivalent to a star formation rate of about $ 1 M_{\sun}yr^{-1}$ at $z \sim 0.4$, with typical errors of about 15\microJy.

\vspace{0.7in}
\section{Data Analysis}

\subsection{Determination of Galaxy Properties}

\subsubsection{Star Formation Rates}

The star formation indicators available to us are the optical [OII]3727\AA, \Hb, and \Ha emission lines, and the \24m mid-infrared flux. In principle, the most direct measure of star formation rates comes from extinction--corrected hydrogen recombination lines (see, e.g. Kennicutt 1998a for a discussion of the general problem). However, our data is inadequate to determine reliable extinction corrections. Of those methods available to us, many studies have established that using the \24m flux is the best choice. A number of empirical calibrations have been made of the correlation of star formation rate with infrared luminosity. Since most of the mid--IR flux comes from warm dust heated by absorbed UV radiation from HII regions (see e.g. Wang \& Heckman 1996), one would expect the best correlation to be between star formation rate and the IR bolometric luminosity. However, that is an observationally difficult quantity, and luminosities in either the IRAS \IRAS25 band or the Spitzer \24m band are more practical measures. Most studies (e.g. Wu \etal 2005, Calzetti \etal 2005, Alonso-Herrero \etal 2006, Rieke \etal 2009) have looked at correlations between bolometric or \24m luminosities and other measures of star formation rates. Among the most sophisticated analyses is that of Rieke \etal (2009), who make use of spectral templates to predict \24m bolometric corrections, as a function of IR luminosity and as a function of redshift.

However, as Perez-Gonzalez \etal (2006) have pointed out, the true expected correlation should be between IR luminosity and the {\em absorbed} rather than total UV luminosity, since only the former heats the dust. Put another way, the correlation should be between the star formation rate and the sum of the IR luminosity and the {\em escaped} UV luminosity. Perez-Gonzales \etal (2006), Calzetti \etal (2007), Kennicutt \etal (2009, hereafter K09) and Calzetti \etal (2010) have all provided calibrations of this relation. All have substituted the easily--observed escaped \Ha flux for the unobserved UV flux in this analysis. This is not strictly correct, because the ratio of absorbed to escaped radiation is much higher in the UV than it is at \Ha. K09 argue that this discrepancy is compensated for by other factors. This is not necessarily true, but the K09 formulation, also used by Calzetti \etal (2010) is more convenient than a more strictly correct analysis, and the best test of its usefulness is the tightness of the correlation between predicted and true star formation rates.

We shall use the relations between \24m and \Ha luminosities and star formation rate given by Equation 17 of Calzetti \etal 2010, but with a slight modification to remove the discontinuities in their formulation at $L(24) = 4\times 10^{42}$ and $L(24) = 5\times 10^{43}$. With this modification, and renormalizing to a Salpeter IMF, we have\vspace{-0.1in}
\hspace{1.5in}
{\renewcommand{\arraystretch}{1.3}
\begin{tabular}{l}
\\
$L(24) < 4 \times 10^{42}$ \\
$\;\;\;\; SFR = 8.1\times 10^{-42}[L(H\alpha) + 0.020L(24)]$\\
\end{tabular}
}
{\renewcommand{\arraystretch}{1.3}
\begin{tabular}{l}
\\
 $4 \times 10^{42} < L(24) \le 5 \times 10^{43}$ \\
 $\;\;\;\;SFR = 8.1 \times 10^{-42}[L(H\alpha)+3 \times 10^{-9}L(24)^{1.16}] $
\end{tabular}
}
{\renewcommand{\arraystretch}{1.3}
\begin{tabular}{lr}
\\
$L(24) > 5 \times 10^{43}$ &\\
$\;\;\;\;  SFR = 2.86\times 10^{-43}L(24)$&\hspace{1.25in} (4)\\
 & \\
\end{tabular}
}
In this and all following equations for star formation rates, luminosities are in units of $erg\,s^{-1}cm^{-2}$, star formation rates is in units of $M_\sun yr^{-1}$, and we assume a mass scale based on a Salpeter IMF.

At higher redshift the \Ha line is often not observable, and we have no measurements of it for about two thirds of our galaxy sample; \OII is the best substitute. Although the relationship between line strength and star formation rate is more straightforward with the Balmer lines than with \OII, the higher--order Balmer lines are both weaker than \OII and also increasingly complicated by underlying stellar absorption lines. Using \OII instead of \Ha in the Calzetti \etal (2010) method should be equally good or better, since the extinction at \OII is closer to that in the UV. Using the data on normal galaxies from Moustakas \& Kennicutt (2006), as tabulated in Table 2 of K09 (hereafter called the K09T2 data set), as well as our own data, we find best consistency between SFR calculated from L(24) and L(\OII) with that calculated from L(24) and L(\Ha), with the following set of relations:
{\renewcommand{\arraystretch}{1.3}
\begin{tabular}{l}
\\
$L(24) < 4 \times 10^{42}$ \\
$\;\;\;\;  SFR = 8.1\times 10^{-42}[1.3L(OII) + 0.020L(24)]$\\
\end{tabular}
}
{\renewcommand{\arraystretch}{1.3}
\begin{tabular}{l}
\\
 $4 \times 10^{42} < L(24) \le 5 \times 10^{43}$ \\
 $\;\;\;\;SFR = 8.1 \times 10^{-42}[1.6L(OII)+3 \times 10^{-9}L(24)^{1.16}] $
\end{tabular}
}
{\renewcommand{\arraystretch}{1.3}
\begin{tabular}{lr}
\\
$L(24) > 5 \times 10^{43}$ &\\
$\;\;\;\;  SFR = 2.86\times 10^{-43}L(24)$ &\hspace{1.15in} (5)\\
 & \\
\end{tabular}
}

The above relations were derived from low--redshift galaxies. For higher redshift objects, observed L(24) will systematically depart from rest frame L(24), so the application of a k--correction is necessary. Unlike the optical case, where one usually has optical colors with which to calculate the continuum slopes needed to obtain the k correction, mid-IR photometry in adjacent bands is not necessarily available. Instead, we use the tabulated k-corrections by Rieke \etal (2009) which have been calculated from models for star--forming galaxies, which predict the continuum shapes and k--corrections as a function of star formation rate.

 If no information on optical emission lines is available we must fall back on an empirical calibration of SFR vs L(24). Figure 4 presents the distribution of the star formation rate calculated from Equations 4 or 5 versus  L(24) for the K09T2, SWIRE and ICBS data sets. The black line represents Equations 4 and 5 in the case where L(\Ha) and L(OII) and zero, expected when galactic extinction is so high that all the UV flux is absorbed and reemitted in the IR.  As it should, this line follows the lower envelope of the galaxy data. The green line represents the best fit to the data, and is of the form\hspace{2.5in}
%\vspace{-0.1in}
 {\renewcommand{\arraystretch}{1.2}
\begin{tabular}{l}
\\
 $L(24) < 2.5 \times 10^{42}$ \\
 $\;\;\;\; log(SFR) = 0.81[log(L24) - 42.40]  $
\end{tabular}
}
{\renewcommand{\arraystretch}{1.2}
\begin{tabular}{lr}
\\
$2.5 \times 10^{42} < L(24) \le 5 \times 10^{43}$ &\\
$\;\;\;\;  log(SFR) = 0.86[log(L24)-42.40]$&\hspace{0.75in} (6)\\
 & \\
\end{tabular}
}
This relation seems to be valid for galaxies at all redshifts and all \24m luminosities within our samples.

%Figure 4
\vspace{0.1in}
\begin{figure}[h]
\figurenum{4}
\epsscale{1.1}
\plotone{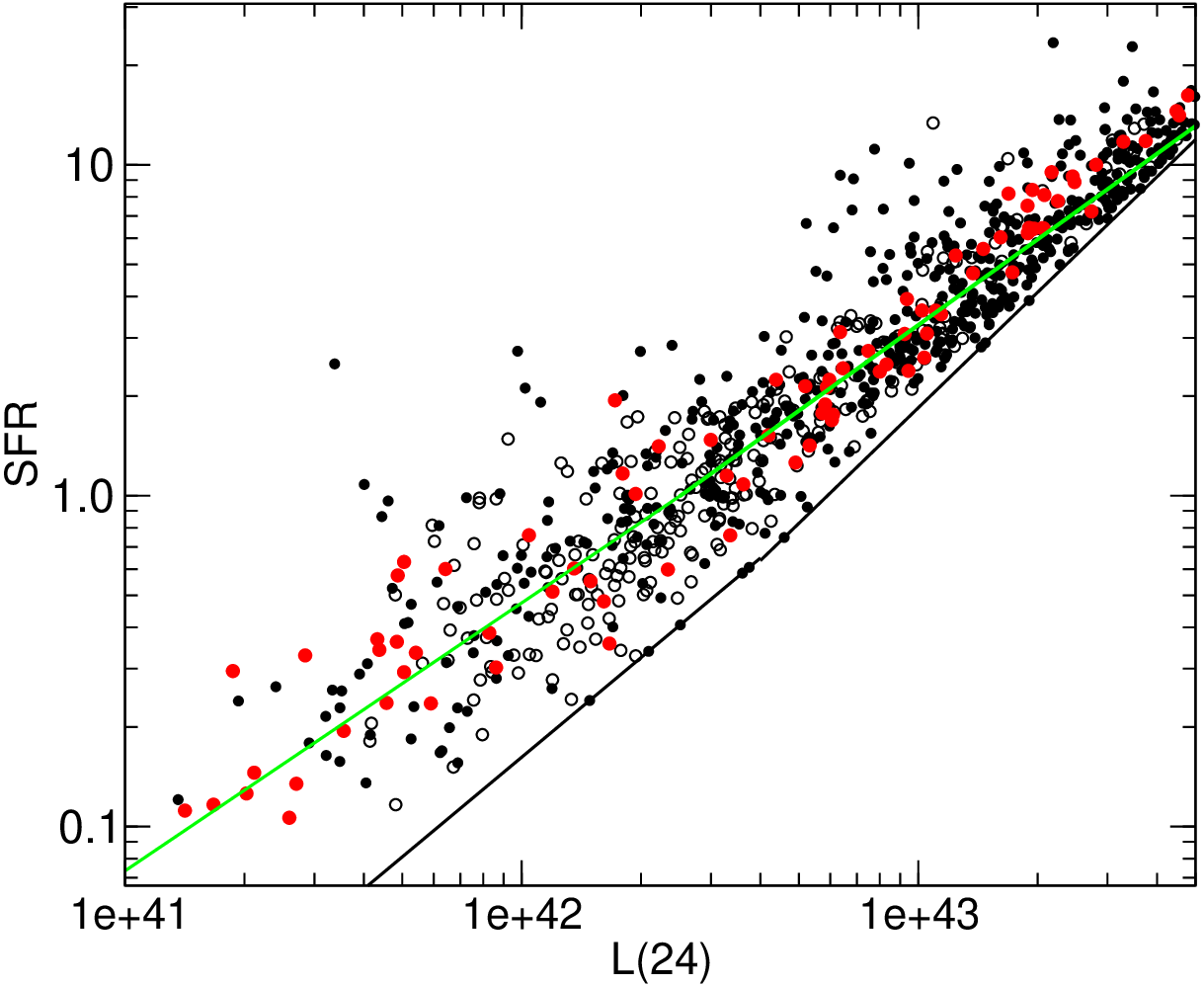}
\figcaption{Star formation rate, calculated from Equations 4 and 5  vs \24m luminosity, for galaxies in the K09T2 (red cirlces), SWIRE (open circles) and ICBS (black circles) samples. The green line is a best fit to the data; the black line represents Equations 4 and 5 with no optical emission contribution}
\end{figure}

If mid--infrared photometry is not available, the only recourse is optical emission lines. In principle, one should be able to use line ratios to correct these lines for extinction, as has been used to obtain, e.g. $L(H\alpha)_{tot}$ for the K09T2 sample. Unfortunately, this requires better data than is usually available for faint galaxies. Experimenting with the ICBS and even the SWIRE data sets demonstrate that using Balmer lines, or Balmer to [OII] ratios to correct the optical line strengths only introduces noise, without improving either random or systematic errors in star formation rates. Instead, we will derive empirical relations between observed line luminosities and star formation rates.

 In Figure 5 we present the relation between star formation rate and {\em observed} \Ha luminosity, for galaxies, in the K09T2 sample, and in the SWIRE and ICBS samples, where star formation rates were calculated using Equations 4 and 5. All three data sets show a similar relation between SFR and L24; the best solution is shown by the green line. The scatter about the line has a dispersion $\sigma(log(SSFR)) \sim 0.16$ for the K09T2 sample, and 0.25 larger for the noisier ICBS data.\hspace{2.5in}
 {\renewcommand{\arraystretch}{1.2}
\begin{tabular}{l}
\\
 $L(H\alpha) < 7\times10^{40}$ \\
 $\;\;\;\; log(SFR)= 1.09(log(L(H\alpha))-40.85) $
\end{tabular}
}
{\renewcommand{\arraystretch}{1.2}
\begin{tabular}{lr}
\\
$L(H\alpha) \ge 7\times10^{40}$ &\\
$\;\;\;\;log(SFR) = 1.31(log(L(H\alpha))-40.85)$&\hspace{0.55in} (7)\\
 & \\
\end{tabular}
}
%Figure 5
\vspace{0.2in}
\begin{figure}[h]
\figurenum{5}
\epsscale{1.1}
\plotone{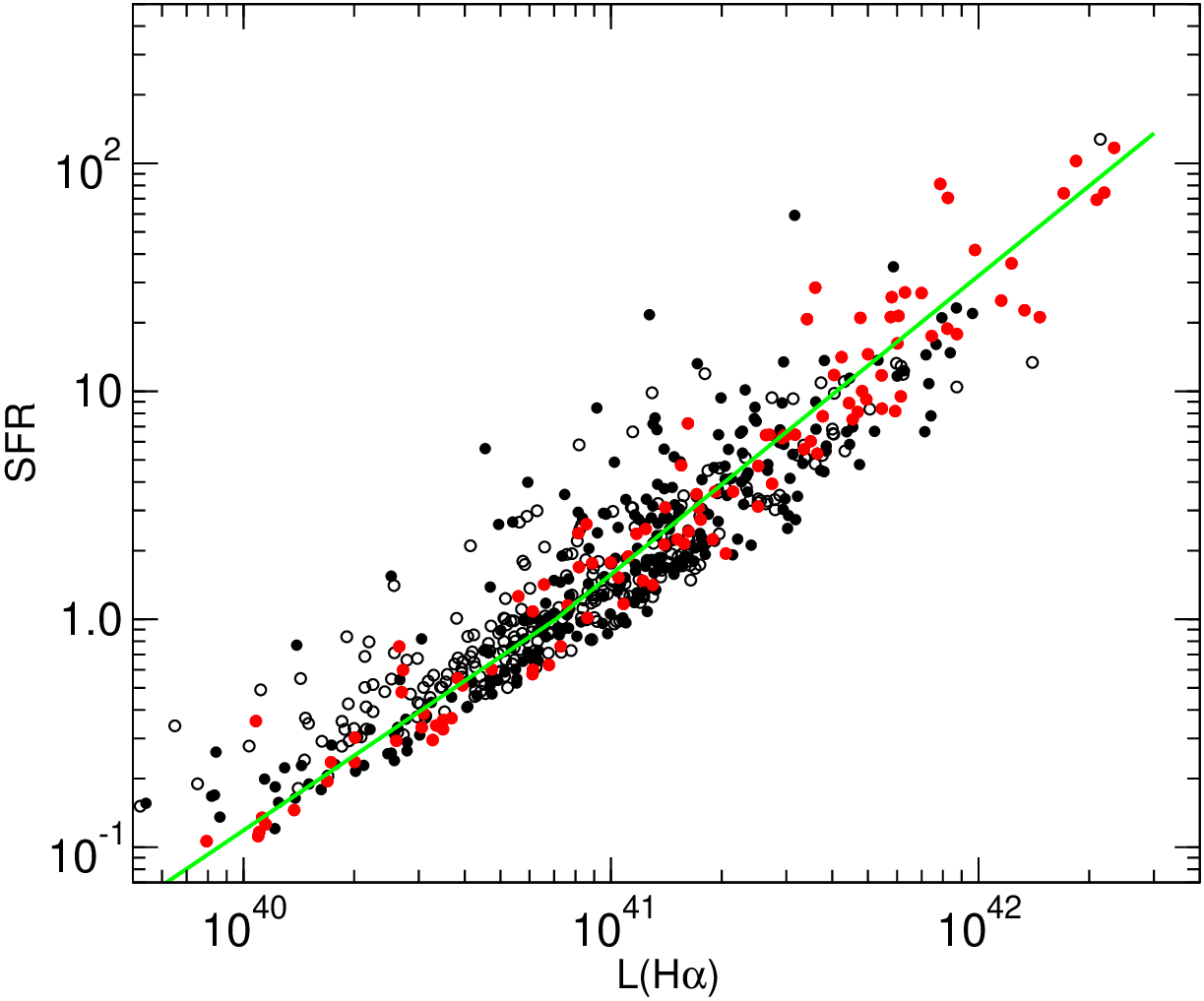}
\figcaption{Total star formation rate vs observed \Ha luminosity for galaxies in the K09T2 sample (red points),the ICBS (black points), and SWIRE samples (open circles). The green line is a best fit to the data.}
\end{figure}

If \Ha is not observable, \Hb is the next best choice. In Figure 6, we present the relation between star formation rate and observed \Hb luminosity, for the same data sets as presented in Figure 5; the best solution for the ICBS data is.
\setcounter{equation}{7}
\begin{equation}
log(SFR)=1.18(log(L(H\beta))-40.18)
\end{equation}

%Figure 6
\vspace{0.25in}
\begin{figure}[h]
\figurenum{6}
\epsscale{1.1}
\plotone{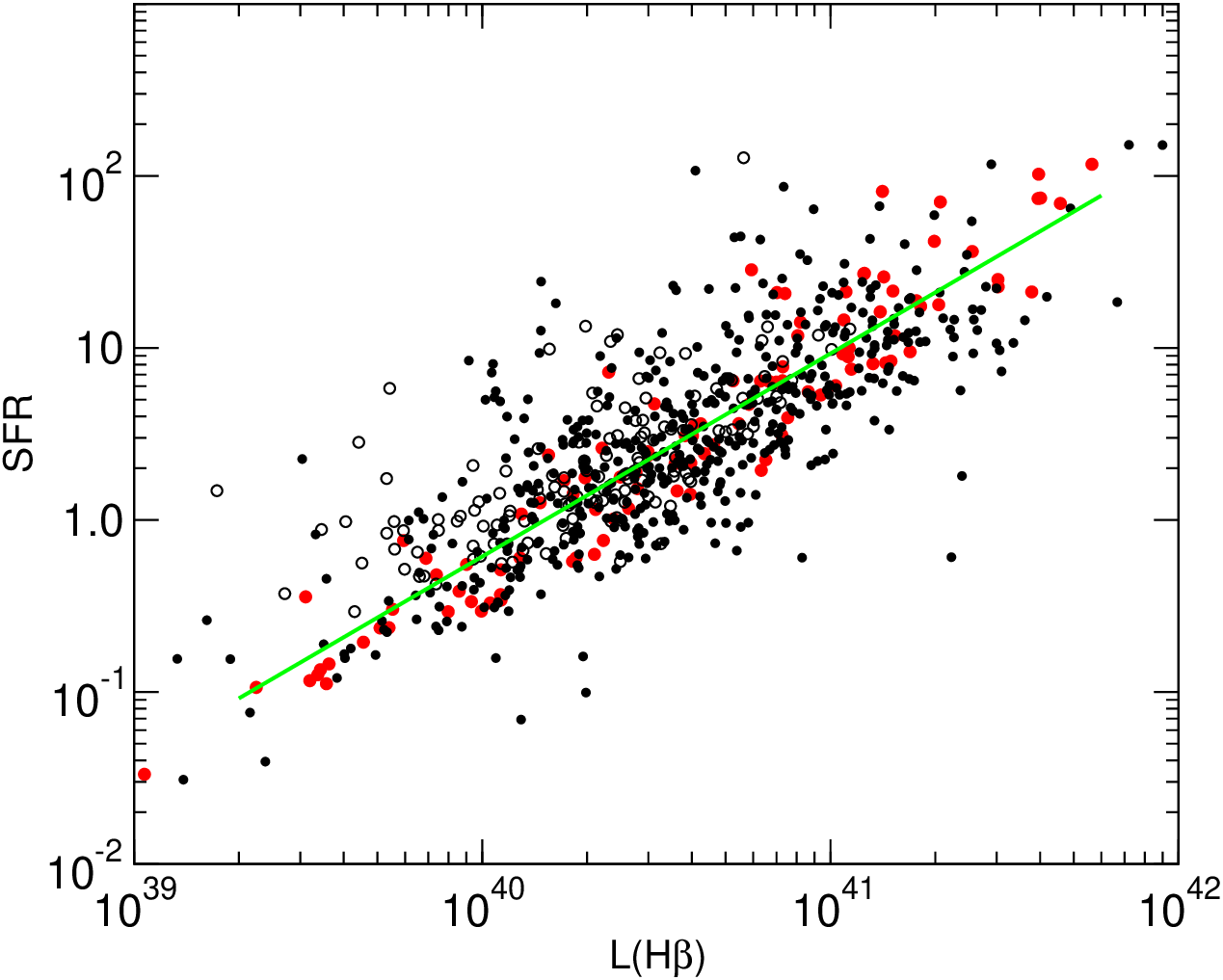}
\figcaption{Total star formation rate vs observed \Hb luminosity for galaxies in the K09T2 sample (red points) and in the ICBS (black points) and SWIRE samples (open circles). The green line is a best fit to the data.}
\end{figure}

The scatter in the \Hb determined SFR is larger:  $\sigma(log(SSFR)) \sim 0.25$ for the K09T2 sample, somewhat larger for the ICBS data. 

Finally, if even \Hb is unobservable, we must fall back on [OII], for which the calibration is presented in Figure 7. The best solution for all the data is

\begin{equation}
log(SFR)=1.10(log(L(OII))-40.39)
\end{equation}

The scatter  for \OII is  $\sigma(log(SSFR)) \sim 0.43$ for all samples.

%Figure 7
\begin{figure}[h]
\figurenum{7}
\vspace{0.3in}
\epsscale{1.1}
\plotone{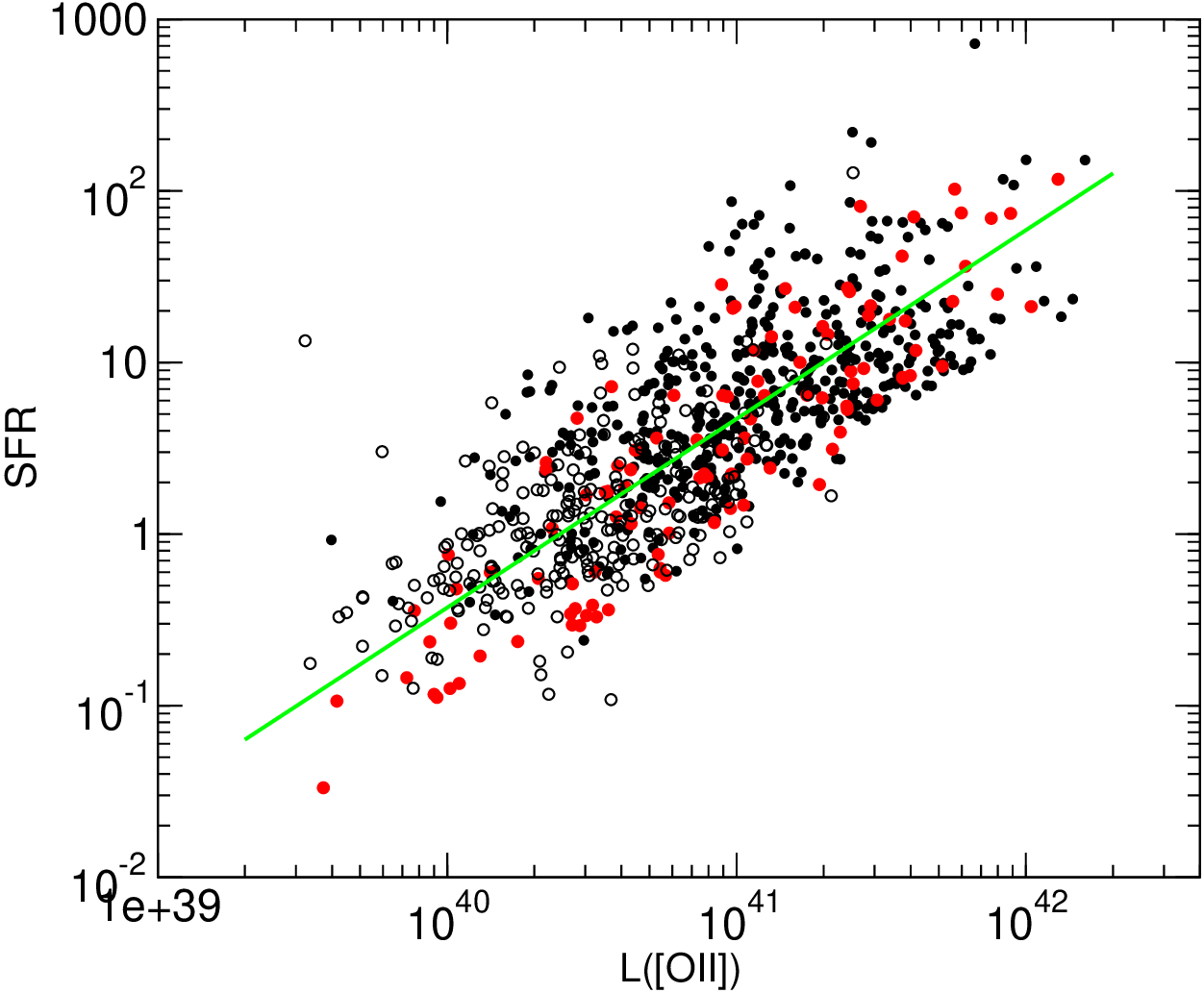}
\figcaption{Total star formation rate vs observed \OII luminosity for galaxies in the K09T2 sample (red points) and in the ICBS and SWIRE samples (black points). The green line is a best fit to the data.}
\end{figure}

It must be emphasized that all of these star formation rate calibrations, and particularly those using only one optical line or IR band, are only claimed to be valid for the star formation rates and galaxy types used in the calibration. However, the galaxy sets which we use should fairly sample normal luminous galaxies are redshifts $z \le 1.0$. The similarity in the relations between, for example, the K09T2 and ICBS galaxy samples, which contain galaxies at quite different redshifts, observed in very different manners, gives us some confidence that these calibrations are indeed appropriate for our sample, and for any of the other galaxy samples produced by surveys of the general galaxy population of the universe. They may or may not be equally applicable to unusual objects such as ULIRG's, or to extreme dwarf galaxies, or to objects at very high redshift.

\subsubsection{Internal Extinction}

%Figure 8
\begin{figure}
\vspace{0.3in}
\figurenum{8}
\epsscale{1.1}
\plotone{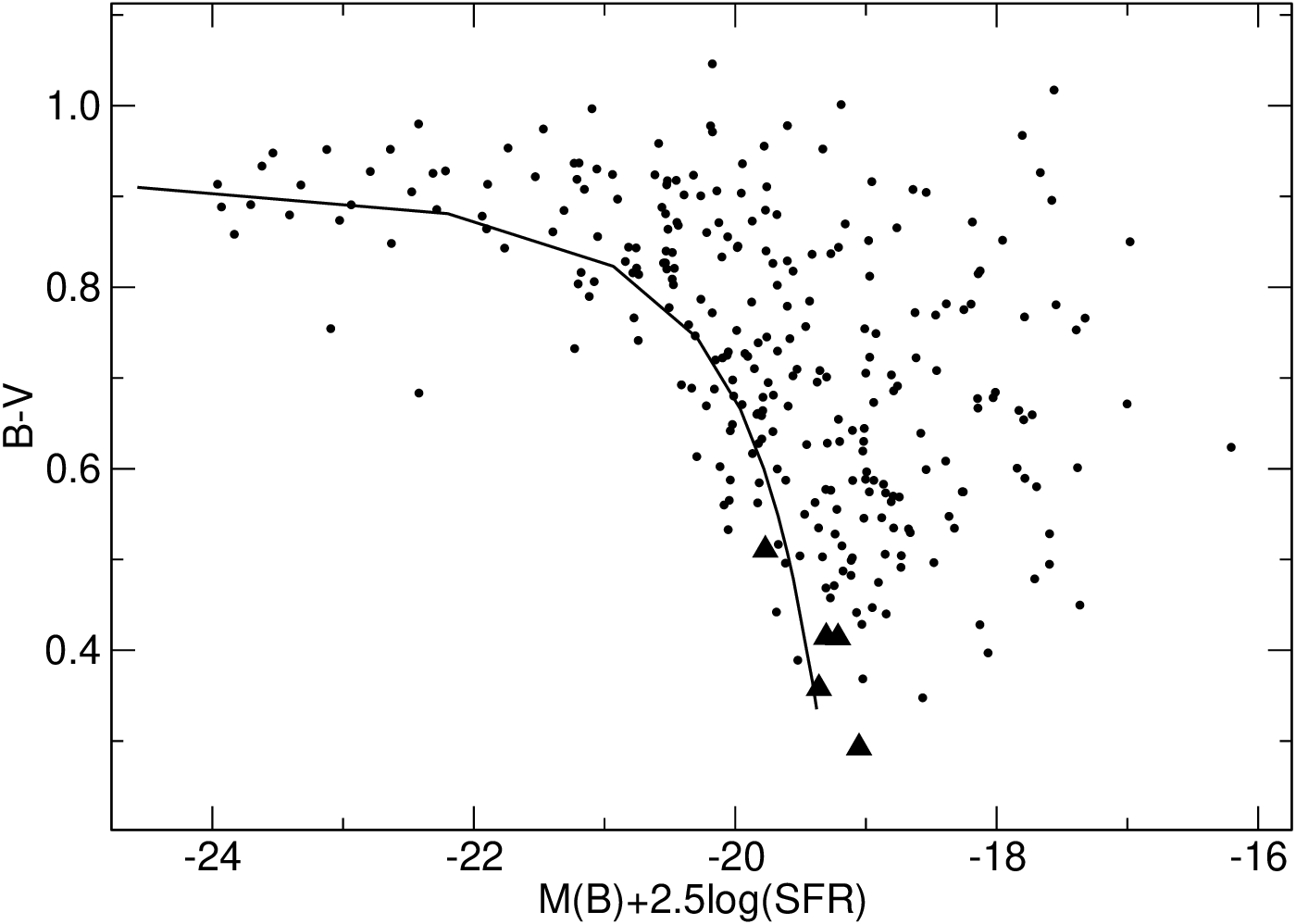}
\figcaption{Observed rest frame B-V colors of galaxies in the SWIRE sample vs the quantity M(B)+2.5log(SFR). Large triangles are objects with $A(V) \le 0.35$. The solid line is the predicted relation for delayed exponential models.}
\end{figure}

For those galaxies with detected \24m flux, and detected \OII or \Ha flux, we can calculate $AV_{em}$, the extinction towards the emission line regions, from the ratio the star formation rate calculated from Equations 4 or 5, and that obtained from the same equations in the limit of $L(24) = 0$, i.e. the case of zero extinction.  However, sometimes it will be useful to know the extinction towards the total stellar continuum of a galaxy. (For example, in Section 3.2 we shall use the dereddened galaxy colors as a starburst criterion). We determine this by the  method described below, using the SWIRE data set. Figure 8 presents the relation, for objects in this  sample,  between the observed rest frame B-V colors and the quantity $S\!F\!R\!M \equiv M_B+2.5log(S\!F\!R)$, i.e. the star formation rate per unit blue luminosity in magnitude form. Objects with emission line extinctions $AV_{em} \le 0.35$ are shown as large triangles. The solid line is the prediction of a set of galaxy models  computed with the Padova evolutionary tracks (Bertelli \etal 1994), adopting a Salpeter IMF with masses in the range 0.15--120 M$_\odot$, and with star formation histories following the form introduced by Gavazzi et al. (2002):

\begin{equation}
SFR \sim \frac{t}{\tau^2} e^{-t^2/2\tau^2}
\end{equation} 

We shall call these {\em delayed exponential} models These models use the observed stellar libraries of Jacoby \etal (1984) in the optical ($\sim 3400$ to 7400 \AA), and they were extended to the ultraviolet and infrared with the theoretical libraries of Kurucz (1993, private communication). They include emission lines formed in HII regions, that were calculated using the photoionisation code {\sc{cloudy}} (Ferland 1996). The nebular component was calculated assuming {\it case B recombination}, electron temperature of $10^4$ K and electron density of $10^2$ cm$^{-3}$. The source of ionizing photons was assumed to have a radius of 15 pc and a mass of $10^4$ M$_\odot$.

Galaxies with low extinction lie close to the line; all others are redder and fainter (more positive values of SFRM) than the line, as would be expected due to extinction of the stellar continuum. We assume that each galaxy is reddened and dimmed by an amount necessary to move it, along a direction parallel to the reddening vector, from a location along the line to its current position. We call the V band extinction of the galactic stellar population,  determined in this way, $AV_*$, and in Figure 9 present, for the objects in Figure 8, the ratio of $A(V)_{em}$ to $AV_*$, vs SFRM. This ratio of the extinction towards the stellar population to that towards emission line regions has been extensively studied by Calzetti, who finds an average value of 0.5 (Calzetti 2001). For those galaxies with well determined ratios (i.e. those with significant values of A(V), there is a remarkably tight relation between $AV_*/AV_{em}$ and star formation rate per {\em observed} luminosity. Galaxies with weak  star formation have low ratios of $AV_*/AV_{em}$, in other words most of the extinction is close to the HII regions, while, in galaxies with vigorous star formation the extinction is spread throughout the galaxy.

%Figure 9
\vspace{0.3in}
\begin{figure}[h]
\epsscale{1.1}
\figurenum{9}
\plotone{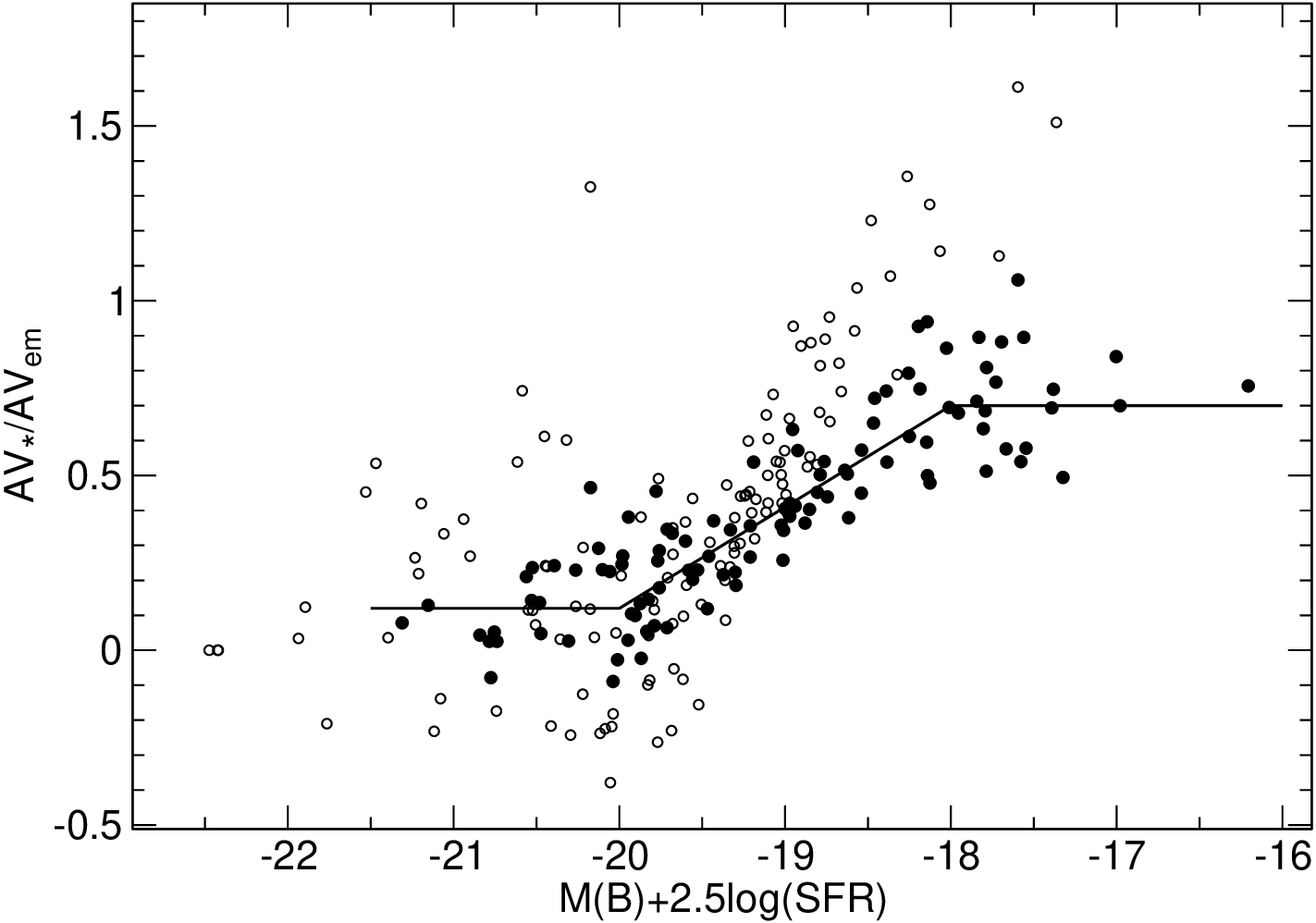}
\figcaption{ratio of $AV_*$ to $AV_{em}$ vs M(B)+2.5log(SFR), for objects in Figure 8. open circles- all objects; filled circles- objects with $AV_{em} \ge 1.0$, which should have more reliable values of $AV_*/AV_{em}$. The solid line is the relation summarized in Eq. 11.}
\end{figure}

 There is a simple explanation of this. Let us make two (oversimplified) assumptions:  (1) all dust is associated with star--forming regions, and (2) the star--forming regions are distributed over the volume of the galaxy in the same way as the stars. Therefore, the average path length for a photon exiting the galaxy will be the same for emission line and continuum photons. Now, if $f$ is the number of star--forming regions along the line of sight of a photon departing the galaxy, then the number of dust clouds encountered by a photon from an HII region within a star--forming region on the way out of the galaxy is $P_{em} \sim 1+f$,  while the number of dust clouds encountered by photons from  a star in the general galaxy population is $P_*\sim f$. Thus, $AV*/AV_{em} = P_*/P_{em} \sim f/(1+f)$, which goes from 0 for small $f$ to unity for large $f$. While an undoubted oversimplification, a qualitatively similar trend must exist as long as some fraction of the galactic dust is associated with star--forming regions, which we know is true.

The solid line in Figure 9 is the relation:
\begin{equation}
\begin{array}{ll}
S\!F\!R\!M\! <\! -20      &        r = 0.12 \\
-20\! <\! S\!F\!R\!M\! <\! -18.1  & r = 0.12 +0.305(S\!F\!R\!M+20) \\
S\!F\!R\!M\! >\! -18.1     &      r = 0.70
\end{array}
\end{equation}

\subsubsection{Mass determination}

%\begin{table*}
%\caption{Polynomial Coefficients for $M/L_B$ vs B-V}
%\begin{tabular}{crrrrrr}
%Redshift &$a_0$ & $a_1$ & $a_2$ & $a_3$ &$a_4$ & $a_5$ \\
%0.00-0.25 & -4.232 & +34.649 & -121.052 & +220.2961 & -197.3101 & +69.0331 \\
%0.25-0.55 & -1.600 & + 6.856 & -9.352 & +5.2178  &   &  \\
%0.55-1.00 & -1.782 & +8.145 & -12.414 & +7.2779   &   &  
%\end{tabular}
%\end{table*}

Bell \& de Jong (2001) present simple prescriptions, based on Bruzual \& Charlot population models, for determining the mass--to--light ratios of galaxies from optical broadband colors. As they point out, their prescription for rest frame B-V colors, for galaxies with a Salpeter IMF and solar metallicity: $log(M/L_B) =-0.51+1.45(B-V)$, has the great virtue that it is almost parallel to the standard reddening vector. Thus, although mass--to--light ratios derived by this method are very sensitive to galactic extinction, {\em masses} are not since they are the product of two quantities, $L_{gal}$ and $(M/L)_{gal}$ with almost exactly inverse dependence of extinction.  If the predictions of such population models are close to correct, our derived masses will be as well.  The most important dependence in such models is on the initial mass function, and we make the same assumption in calculating masses- a Salpeter IMF, as we have made in calibrating star formation rates.

%Table 2
\begin{deluxetable}{rrrrrrr}
\tablecaption{Polynomial Coefficients for $M/L_B$ vs B-V}
%\tablewidth{0pt}
\tablehead{\colhead{Redshift} & \colhead{$a_0$} & \colhead{$a_1$} & \colhead{$a_2$} & \colhead{$a_3$} & \colhead{$a_4$} & \colhead{$a_5$} }
\startdata
0.00-0.25 & -4.23 & 34.64 & -121.05 & 220.29 & -197.31 & 69.03 \\
0.25-0.55 & -1.60 & 6.85 & -9.35 & 5.22  &   &  \\
0.55-1.00 & -1.78 & 8.14 & -12.41 & 7.28   &   &  
\enddata
\end{deluxetable}

We use a variant of the Bell \& de Jong approach, but calculate mass--to--light ratios using the delayed exponential models described in the previous section. We calculate the predictions at the epochs observed at  redshifts between 0.0 and 1.1; the results are presented, in comparison with the Bell \& de Jong prescription, in Figure 10. 

%Figure 10
\begin{figure}[h]
\vspace{0.2in}
\figurenum{10}
\epsscale{1.0}
\plotone{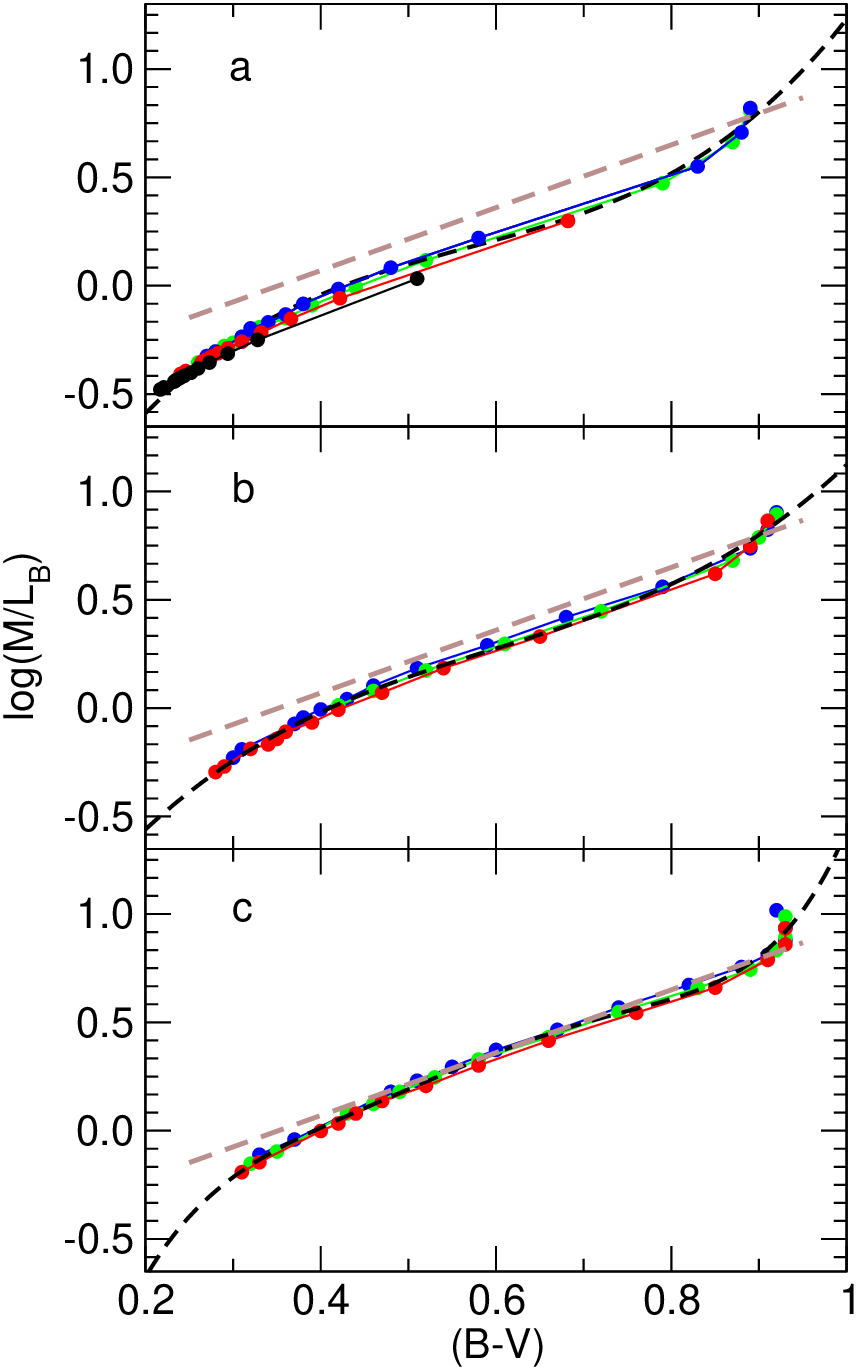}
\figcaption{B band mass--to--light ratios vs (B-V) color. The brown line represents the Bell \& deJong (2001) solution $log(M/L_B) =-0.51+1.45(B-V)$. The dashed black lines are fits to the  delayed exponential model predictions, shown by colored points and lines,  in each redshift range. a- redshifts of 0.6 (blue), 0.7 (green), 0.9 (red), and 1.1 (black). the fit is to only the 0.6 and 0.7 relations. b- redshifts of 0.3 (blue), 0.4 (green), and 0.5 (red). c- redshifts of 0.0 (blue), 0.1 (green), and 0.2 (red).}
\end{figure}

Our results are similar to those of Bell \& de Jong, but they vary with epoch, and a linear relation is not the best representation, Instead, we use the polynomials of the form $\log (M/{L_B}) = \sum\limits_0^n {{a_i}(B - } V{)^i}$, with values of $a_i$ presented in Table 2.

These polynomials do not reflect the full upturn seen in the models at the red end. This is deliberate; given errors in the observed colors, a relation as steep as the model curves would produce much too large values of M/L for some galaxies. Since these relations are not exactly parallel to the reddening vector, we correct both M/L and $L_{gal}$\ for extinction before calculating masses. If a continuum extinction values is not available for a galaxy, we assume a value $A(V)_* = 0.4$. Because the above relations are close to the reddening line, the effect of incorrect extinction values on the derived masses will be small.

\subsection{Starburst Criteria}

The role that starbursts might play in the evolution of galaxy populations, both in clusters and the field, has been a subject of considerable attention and controversy, at least since Dressler \& Gunn (1983). Papers II and III of this series (Dressler \etal 2013, Oemler \etal 2013) will examine  the evidence provided by this survey in some detail. The ICBS data provide multiple means of detecting starbursts.  

We first reexamine the usual spectroscopic indicators. The equivalent widths of [OII]3727\AA\   and \Hd have been used for many years as starburst indicators, (e.g. Dressler \& Gunn 1983, Couch \& Sharples 1987, Dressler \etal 1999); too strong an [OII] line can only be produced in galaxies during a starburst, and too strong an \Hd line is produced either during or after a starburst. 

We wish to recalibrate the threshold strengths of both lines which separate normal from bursting star formation, using stellar population models and empirical evidence. The behavior of \OII is fairly simple to understand. Its equivalent width is the ratio of the emission produced by HII regions to the nearby stellar continuum, to which stars of all ages contribute, but younger, hot stars contribute the most. Dust extinction can only produce one possible effect: selectively diminishing the HII emission line strength relative to that of the more broadly distributed blue stars. Figure 11 presents the EW([OII]) vs SSFR distribution of galaxies in the SWIRE and K09T2 samples, and of ICBS galaxies. The red lines are the predictions of the extinction--free delayed exponential models for normal galaxies at redshifts (bottom to top) of 0.0, 0.5, and 1.0. These predictions can be taken as maximum allowed values of EW([OII]), since selective extinction of star--forming regions, as described in Section 3.1  will generally reduce the observed value of EW([OII]) below its extinction--free value. We take as the threshold for starbursts the green line, which has the form $EW([OII]) = \sum\limits_0^n {{a_i}{{(\log (SSFR))}^i}}$, with values of $a_i$ as presented in Table 3. Necessarily an OII criterion for starbursts must miss a significant fraction of them, since objects which start off far from the green line may not cross the threshold during even a strong starburst.

%Figure 11
\begin{figure}[h]
\vspace{0.2in}
\figurenum{11}
\epsscale{1.15}
\plotone{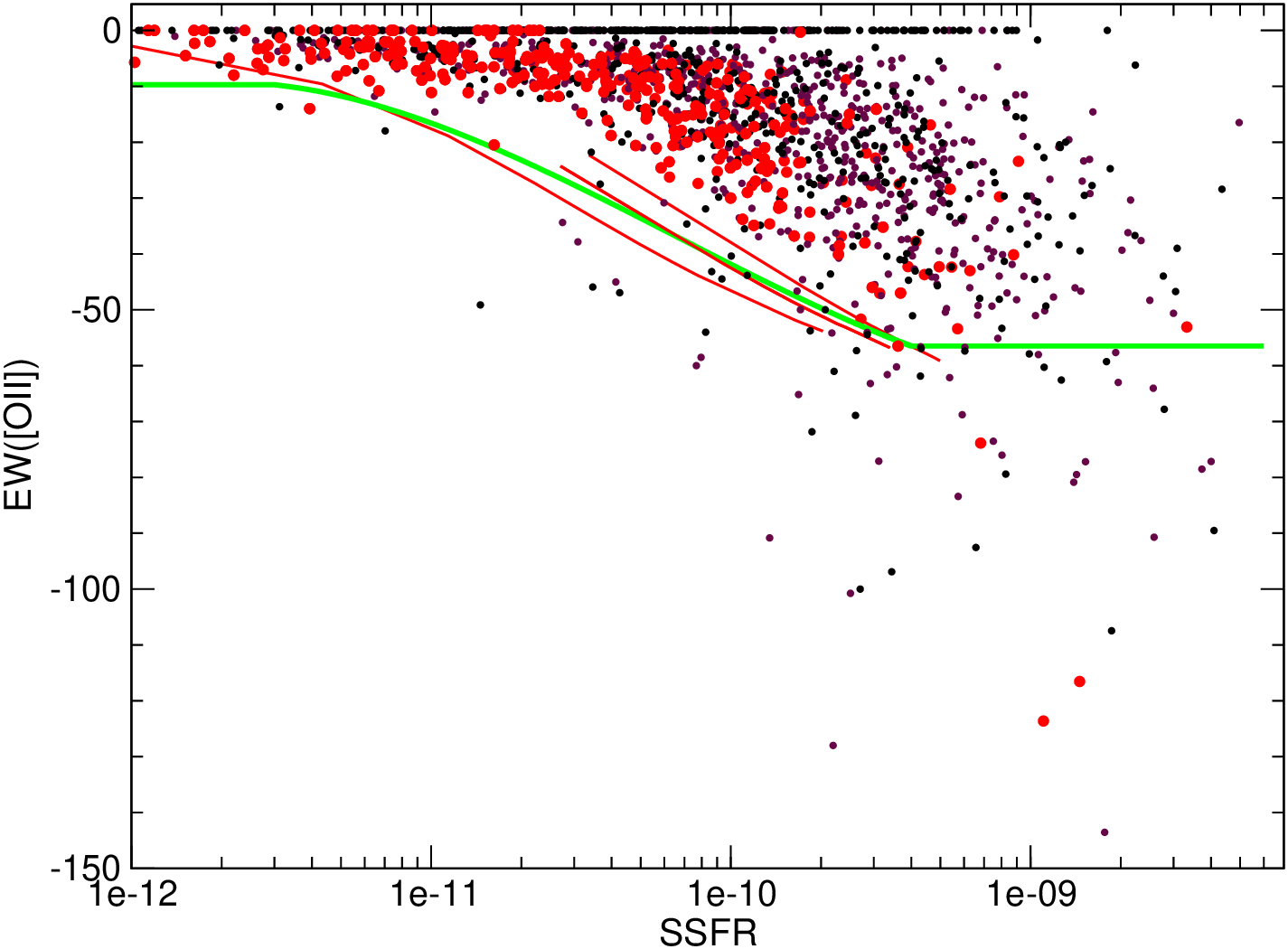}
\figcaption{EW([OII]) versus specific star formation rate. Red points: galaxies from SWIRE and K09T2, black points- ICBS galaxies. Red lines are the relations predicted by our models for redshifts of (top to bottom) 0,0, 0.5, and 1.0. Green line is the adapted threshold for starbursts.}
\end{figure}

The effect of extinction on the \Hd line is more complex, and has been studied in some detail by Poggianti \etal (2001). In a dust--free system, \Hd will be strongest in systems dominated by A stars- such as post--starburst galaxies in which the OB stars have died but the A stars have not. In systems dominated by older, cooler stars the line is weaker, in younger, hotter systems the line begins to be filled in by HII region emission. The effect of adding dust will depend on the relative distribution of dust, HII regions, A stars, and cool stars, and may either enhance or diminish the strength of the \Hd line. Some starbursts are known with no visible optical emission lines but strong \Hd (Smail \etal 1999, Dressler \etal 2009b), presumably cases where the HII regions are heavily absorbed, but the A star products of the starburst have migrated away from the dustiest regions. 

Because of the complexities, a theoretical prediction of \Hd strength is impossible for any but dust--free systems. For galaxies with significant Population I, an empirical determination is necessary.  \Hd correlates equally well with \OII strength and with broadband color; we choose, following previous practice, to use \OII. In Figure 12 we present the dependence of \Hd on EW([OII]) for galaxies in the SWIRE and K09T2 samples.  Because there is no reliable theoretical prediction, and because the distribution is broad, we define a measure \DHd, which is the strength of \Hd relative to the line defined in Figure 12, with values, $\EWOII <=3.0 \AA:    EW(\Hd) = 3.0 $; $\EWOII > 10\AA:     EW(\Hd) = 5.5$, and a linear increase between the two.

%Table 3
\begin{deluxetable}{crrrr}
\tablecaption{Polynomial Coefficients for EW([OII]) vs SSFR}
\tablewidth{0pt}
\tablehead{\colhead{SSFR} & \colhead{$a_0$} & \colhead{$a_1$} & \colhead{$a_2$} & \colhead{$a_3$} }
\startdata
 $< 3\times 10^{-12}$  &   -10 & & & \\
$3\times 10^{-12}- 4\times 10^{-10}$ & 3940 & 1126 & 123.2 & 4.033 \\
$\ge 4 \times 10^{-10}$ &   -56 & & &
\enddata
\end{deluxetable}

The quantity \DHd is a measure of  the likelihood that an object is a starburst; within the SWIRE sample about 95\% of objects have values of \DHd less than zero. As with our \OII measure, the \DHd criterion will necessarily miss some fraction of starbursts in galaxies whose initial location in the \OII-\Hd plane is far from the $\DHd=0$ line. However, unlike our OII measure, there will be, at any positive value of \DHd, some chance that the object is {\em not} a starburst, but merely an outlier in the \OII-\Hd distribution.  For dust--free passive galaxies we can have more confidence; theoretical models (Poggianti \etal 1999) agree with empirical evidence that 3\AA\  is an upper limit to the \Hd strength of normal galaxies.

%Figure 12
\begin{figure}[h]
\vspace{0.35in}
\figurenum{12}
\epsscale{1.1}
\plotone{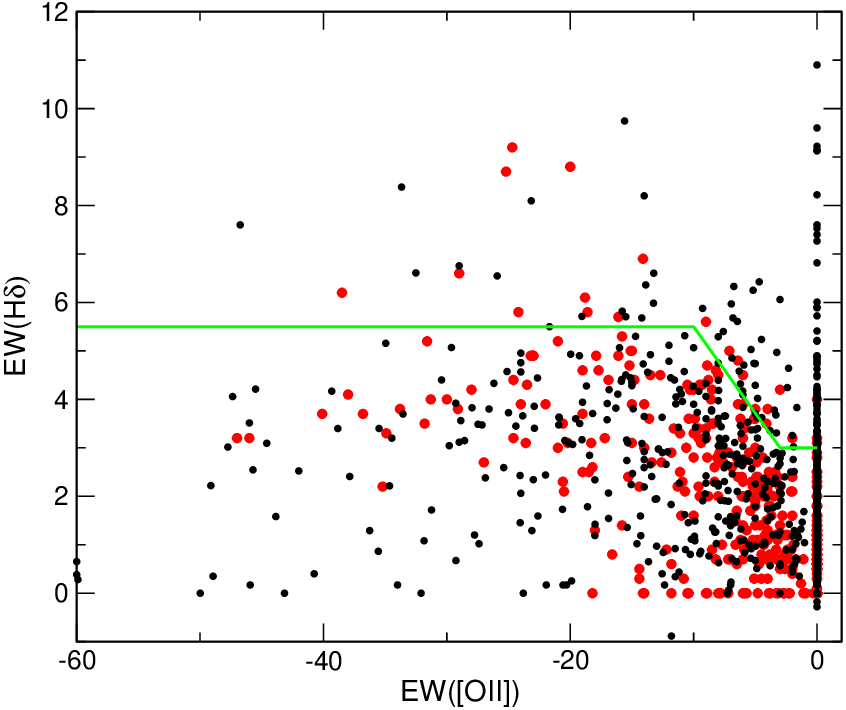}
\figcaption{a- EW(\Hd) vs \EWOII for galaxies in the SWIRE (red points) and ICBS (black points) samples. The green line defines $\DHd=0$. }
\end{figure}

Larson \& Tinsley (1978) were the first to demonstrate that the color distribution of starburst galaxies is broader than that of normal objects.  In Figure 13 we plot the distribution of  rest frame B-V colors, corrected for extinction as described in Section 3.1.2  vs SSFR for galaxies in the MK06 and SWIRE samples, and, as a green line, our delayed exponential model predictions for a redshift of 0.0. (At higher redshifts, the predictions for high values of SSFR move parallel to the $z=0.0$ locus, extending the line to higher SSFR's and bluer colors.)  We have determined morphological classifications for  as many as possible MK06 galaxies, using images from the NASA Extragalactic Database, and have identified AGN's using the line strength criteria of  Kauffmann et al (2003). Galaxies which are morphologically and spectroscopically normal are displayed as large black circles, AGN's and morphologically peculiar galaxies are displayed as red circles. SWIRE galaxies, for which no morphology information is available, are shown as small blue circles.

Normal galaxies have a very tight distribution about the predicted relation, with $\sigma(B-V)_{corr} = 0.035$. On the other hand,  as Larson \& Tinsley originally demonstrated, peculiar galaxies- almost all of which have morphological peculiarities suggestive of interactions, and therefore of starbursts,  scatter more widely. The two black lines, containing the region of normal galaxies, have the following form, where $S = log(SSFR)+11$:

\begin{equation}
\begin{array}{ll}
S<0:  &   (B\!-\!V)_{min} =  0.69-0.130S \\
&  (B\!-\!V)_{max} = 0.89-0.130S \\
& \\
S\ge 0: &  (B\!-\!V)_{min} =  0.69-0.315S \\
&  (B\!-\!V)_{max}=   0.89-0.315S
\end{array}
\end{equation}

%Figure 13
\begin{figure}[h]
\figurenum{13}
\vspace{0.2in}
\epsscale{1.1}
\plotone{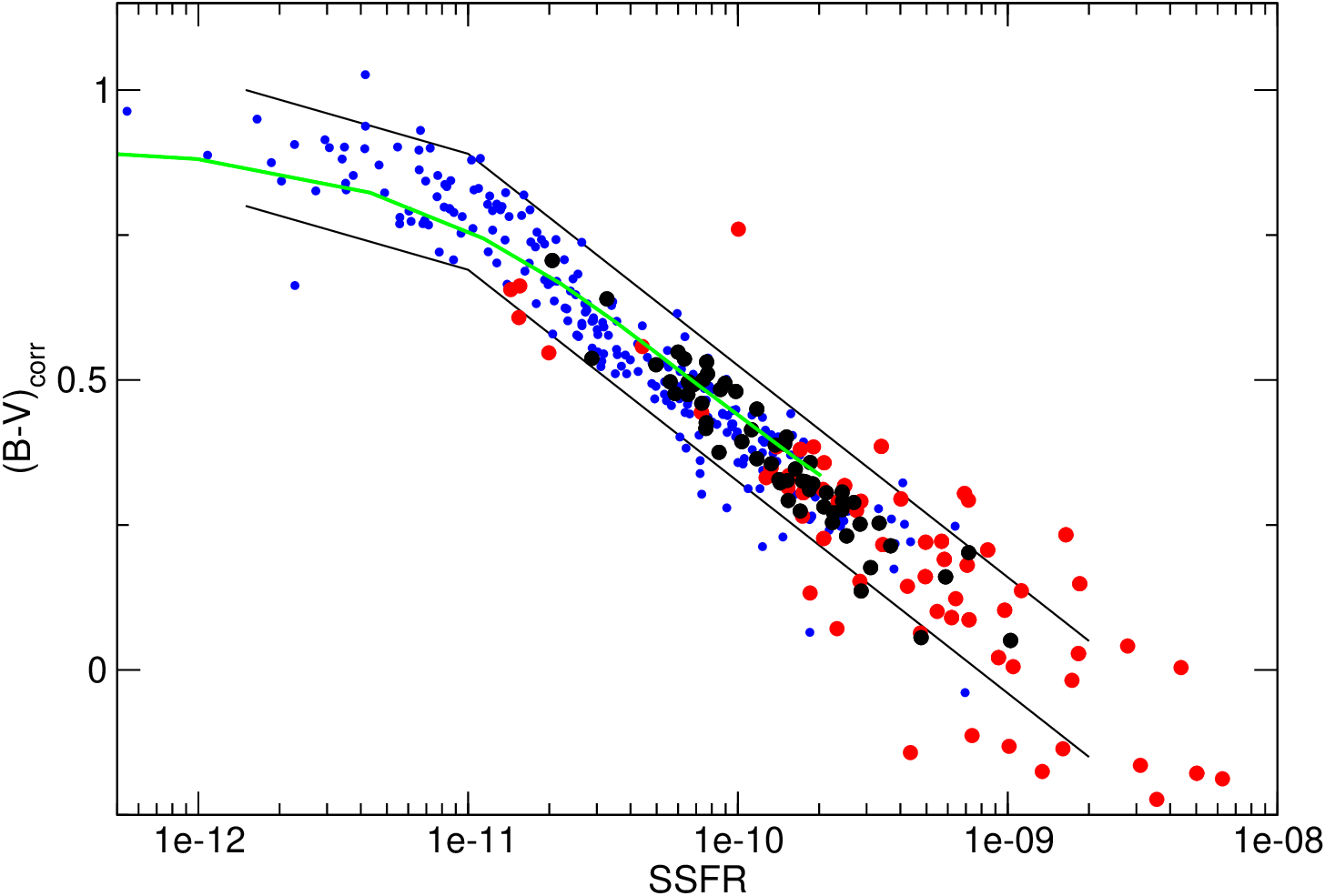}
\figcaption{The distribution of reddening--corrected  rest frame B-V colors of galaxies in the MK06 and SWIRE samples, versus specific star formation rate. green line- predicted relation. Blue points- SWIRE galaxies; black points- normal MK06 galaxies, red points- MK06 galaxies with either morphological peculiarities or spectral indicators of AGN activity. The black lines are our chosen limits for normal galaxies}
\end{figure}

Objects beyond the region defined by these lines are presumed to be in some phase of a starburst. Objects above the top line, with star formation rates too high for their colors, are likely young starbursts. Those below the bottom line, with colors too blue for their star formation rates, are probably young post--starbursts. Only about 1/3 of the morphologically--defined starburst candidates lie beyond the normal region. As with the previously defined OII and \Hd starburst indicators, color selection can only discover a fraction of starbursts. Since most of the outliers have very high specific star formation rates, it will be predominantly the stronger starbursts which are detected by this criterion.

 \subsection{Cluster Properties}
 
 %Figure 14
\begin{figure}[h]
\figurenum{14}
\vspace{0.45in}
\epsscale{1.0}
\plotone{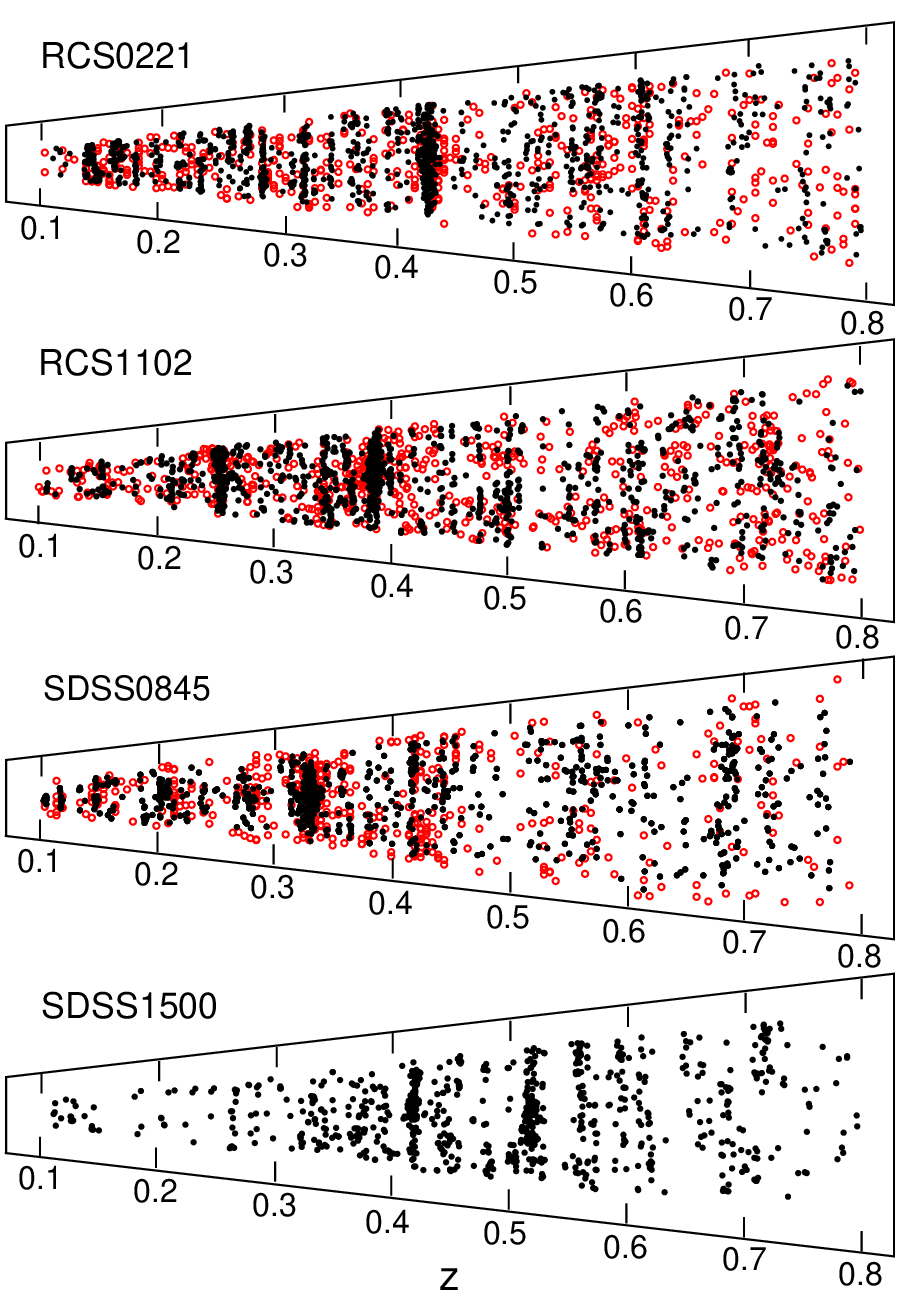}
\caption{Redshift--declination pie diagrams for each of the 4 fields. Objects with redshifts determined by grism specroscopy are shown as black points; objects with only LDP spectroscopy are shown as open red circles. }
\end{figure}
Figure 14 presents redshift-declination pie diagrams for galaxies in the 4 fields in the redshift interval 0.10 to 0.80.  The RCS0221 and SDSS0845 fields each contain one dominant cluster, at redshifts near where the Red Cluster Sequence cluster detection method predicted one to lie. The situation in the other two fields is more complicated; in both there exist two clusters of comparable richness, and with redshifts similar enough that both probably contributed to the RCS detection signal. Pie diagrams in the redshift region of each of the 6 major clusters are presented in Figure 15. Filled circles represent objects that we identify as belonging to the clusters. There is, of course, some ambiguity about the correct redshift cut in some of these clusters; we have chosen to err on the side being too inclusive, so as not to miss any potential cluster members in our later analysis, and so as not to contaminate our field sample by cluster members.

%Table 4
\begin{deluxetable}{lcc@{\hspace{1pt}}c@{\hspace{1pt}}c@{\hspace{1pt}}c@{\hspace{3pt}}c}
\tablecaption{Properties of Clusters}
%\tablewidth{0pt}
\tablehead{\colhead{Cluster}  & \colhead{$N_{gal}$} & \colhead{z}  & \colhead{$r_{200}$(Mpc)} &  \colhead{$L/L*$} & \colhead{$\sigma_{r200}$} & \colhead{$\sigma_{tot}$ } }
\startdata
RCS0221 & 245 (337) & 0.431 & 1.27 & 158 &  941 & 895\\
RCS1102A & 156 & 0.255 & ---  &--- & ---&--- \\
RCS1102B & 245 & 0.386 & 1.39  & 224 & 772 & 697\\
SDSS0845 & 278 (383) & 0.330 & 1.43  & 202 & 1087\tablenotemark{a} & 1031\tablenotemark{a}\\
SDSS1500A & 113 & 0.420 & 1.17 &82 & 798  & 639 \\
SDSS1500B & 160& 0.518 & 1.23  & 182 & 844\tablenotemark{b} &904\tablenotemark{b}
\enddata
\tablenotetext{a}{only includes redshift range 0.317--0.343}
\tablenotetext{b}{ only includes redshift range  0.507--0.531}
\end{deluxetable}

With the cluster membership as defined, Table 4 presents a summary of cluster properties. Numbers in parentheses include LDP spectra. The radius $r_{200}$ is defined in the usual way (Carlberg, Yee, \& Ellingson 1997), $\sigma$ is the velocity dispersion of all cluster members within a projected radius of $r_{200}$, $N_{gal}$ is the observed number of members, and $L/L*$ is  the total cluster luminosity, in units of L*, calculated assuming a Schechter luminosity function with the parameters, $\alpha=-1.05$ and $M_B* = f(z) $ as we determine in Paper III, and correcting for sample incompleteness as a function of magnitude and position. The quantity $L/L*$ should not be over-interpreted: it refers to a volume large compared to either the virial  radius or the typical radius to which clusters are normally observed, and therefore includes much supercluster population. Values are not given for some properties of RCS1102A: the cluster center apparently lies close to the edge of the survey area, and we cannot know what fraction of the total cluster is observed.

%Figure 15
\begin{figure}[h]
\figurenum{15}
\vspace{0.4in}
\epsscale{1.0}
\plotone{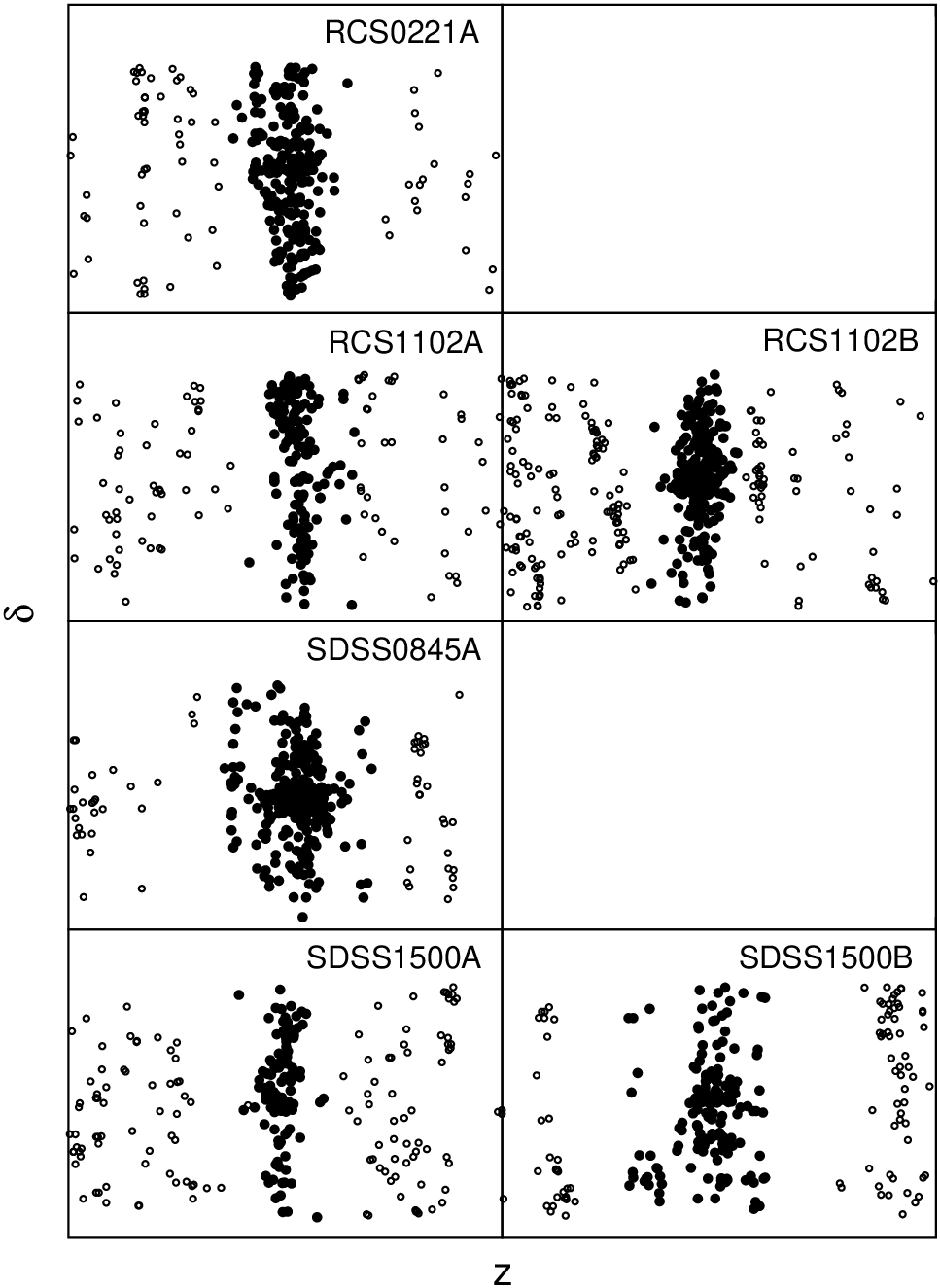}
\caption{ Redshift--declination pie diagrams centered on each of the 6 massive clusters found in the 4 fields. Objects with only LDP redshifts are not included. Open circles are field galaxies; filled circles are galaxies assigned to the clusters. The redshift range in each plot is $z_{cluster} \pm 0.07$, the declination range is 0.7\degr.}
\end{figure}

\section{Summary}

We have obtained photometry and usable spectra of 6002 galaxies in 4 fields of 30\arcmin\ diameter,  from which were measured  absolute magnitudes, rest frame colors, redshifts and absorption and emission line strengths. The 6002 galaxies includes 1394 members of 5 clusters. Deep \24m Spitzer photometry was also obtained for two of the 4 fields. Using new calibrations of star formation rates from optical and IR indicators, star formation rates, or an upper limit were obtained for 96\% of the  galaxies in the redshift sample (71\% detected SFR, 25\% upper limits). From colors, masses were determined for 87\% of galaxies with $z \le 0.7$. In addition, at least one measure of the presence or absence of an ongoing or recent starburst was obtained for 69\% of the redshift sample. 

With these data in hand, we shall examine the evolution of the star formation properties of galaxies in the immediately following papers. Paper II (Dressler \etal 2013) begins the examination of the processes driving the evolution of galaxies infalling into clusters. In Paper III (Oemler \etal 2013) we  examine star formation in field galaxies out to $z=0.6$. Paper IV (Gladders \etal 2013) constructs a more detailed model of field galaxy evolution, using the data discussed in Paper III as well as the observed evolution of the star formation rate density. In forthcoming papers we will elaborate on this model, as well as examine the effect of mergers and interactions on star formation and evolution, the near infrared properties of our galaxy sample, as well as other aspects of the cluster environment.

\section{Acknowledgements}

Oemler and Dressler acknowledge support of the NSF grant AST-0407343. This work is based in part on observations made with the Spitzer Space Telescope, which is operated by the Jet Propulsion Laboratory, California Institute of Technology under a contract with NASA. Support for this work was provided to Dressler, Oemler, Rigby, Bai, and Rieke by NASA through an award issued by JPL/Caltech. Support for this work was provided by NASA to Rigby through the Spitzer Space Telescope Fellowship Program, through a contract issued by the Jet Propulsion Laboratory, California Institute of Technology under a contract with NASA. Vulcani and Poggianti acknowledge financial support from ASI contract I/016/07/0 
and ASI-INAF I/009/10/0. This research has made use of the NASA/IPAC Extragalactic Database (NED) which is operated by the Jet Propulsion Laboratory, California Institute of Technology, under contract with the National Aeronautics and Space Administration.  We have made extensive use of Ned Wright's Online Cosmological Calculator (Wright 2006).

\clearpage

\end{document}